\documentclass[preprint,10pt]{aastex} 

\shorttitle{Large Flaring Loops in PMS Stars}
\shortauthors{Aarnio, Stassun, \& Matt}
\usepackage{lscape,enumerate}

\newcommand{\catwone}{six}
\newcommand{\catwcone}{Six}
\newcommand{\catpone}{25\%} 

\newcommand{\catwtwo}{four}
\newcommand{\catthree}{14}
\newcommand{\catwthree}{fourteen}
\newcommand{\catwcthree}{Fourteen}
\newcommand{\catpthree}{58\%} 

\newcommand{\catwfour}{eight}

\newcommand{\dik}{$\Delta K_{\rm S}$}
\newcommand{\duv}{$\Delta (U-V)$}
\newcommand{\goodguys}{24}

\begin{document}

\title{A Search for Star-Disk Interaction Among the Strongest X-ray Flaring Stars in the Orion Nebula Cluster}

\author{Alicia N.\ Aarnio\altaffilmark{1}, Keivan G.\
Stassun\altaffilmark{1,2}, and Sean P.\ Matt\altaffilmark{3}}

\altaffiltext{1}{Department of Physics and Astronomy, Vanderbilt University, VU Station B 1807, Nashville, TN 37235, USA}
\altaffiltext{2}{Department of Physics, Fisk University, 1000 17th Avenue N., Nashville, TN 37208}
\altaffiltext{3}{NASA Ames Research Center, Moffett Field, CA 94035}

\begin{abstract}
The Chandra Orion Ultradeep Project observed hundreds of young, low-mass stars 
undergoing highly energetic {\sc x}-ray flare events. The 32 most powerful 
cases have been modeled by \citet{Favata:2005} with the result that the magnetic 
structures responsible for these flares can be many stellar radii in extent. 
In this paper, we model the observed spectral energy distributions of
these 32 stars in order to determine, in detail for each star, whether
there is circumstellar disk material situated in sufficient proximity to
the stellar surface for interaction with the large magnetic loops inferred 
from the observed X-ray flares. Our spectral energy distributions span 
the wavelength range 0.3--8~$\mu$m (plus 24~$\mu$m for some stars), allowing us 
to constrain the presence of dusty circumstellar material out to 
$\gtrsim 10$~{\sc au} from the stellar surface in most cases. For \goodguys\ of 
the 32 stars in our sample the available data are sufficient to constrain 
the location of the inner edge of the dusty disks.  \catwcone\ of these (\catpone) 
have spectral energy distributions consistent with inner disks within reach of 
the observed magnetic loops. Another \catwtwo\ stars may have gas disks interior 
to the dust disk and extending within reach of the magnetic loops, but we cannot 
confirm this with the available data. The remaining \catthree\ stars (\catpthree) 
appear to have no significant disk material within reach of the large flaring loops.
Thus, up to $\sim 40$\% of the sample stars exhibit energetic {\sc x}-ray flares 
that possibly arise from a magnetic star-disk interaction, and the remainder are 
evidently associated with extremely large, free-standing magnetic loops anchored 
only to the stellar surface.  
\end{abstract}


\section{INTRODUCTION\label{intro}}

The recent large {\sc x}-ray surveys of the Orion and Taurus star-forming
regions performed by {\it Chandra} and {\it XMM} \citep[i.e.,\ {\sc
coup}, {\sc xest}:][]{COUP:2005,XEST:2007} provide an unparalleled
opportunity to study the magnetic activity of young, low-mass stars.
These deep observations spanning long temporal baselines (e.g.,\
the {\sc coup} {\sc x}-ray light curves span 13 days with near-continuous
time coverage) reveal that low-mass pre-main-sequence ({\sc pms}) stars
possess {\sc x}-ray luminosities 3--4 magnitudes greater than that of the
present-day Sun and exhibit extremely energetic flaring events with
high frequency.

Detailed analyses of these flares reveal that they are similar to
solar flares, but are orders of magnitude more energetic and larger in
physical size. In particular, \citet{Favata:2005} subjected the 32 most
energetic flares observed by {\sc coup} to analysis via a standard
uniform cooling loop model 
\citep{Reale:1997,Sylwester:1993,Priest:2002,Favata:2003},
with which they derived the properties of the 
magnetic coronal loops that participate in the flare events. They found 
that these magnetic loops were extremely large---extending tens of 
stellar radii in some cases---much larger than ever observed on older stars.
Such large-scale flares could have important ramifications for a
number of issues, such as the shedding of stellar angular momentum and mass,
powering of outflows, and ionization/dissipation of circumstellar disks. 

Magnetic loops with sizes on the order of $\sim10$ stellar radii have long been 
postulated as part of magnetospheric accretion scenarios. In this paradigm, a 
large-scale stellar magnetic field threads the inner edge of a circumstellar 
disk, channeling accretion from disk to star 
\citep[e.g.,][]{Camenzind:1990,Konigl:1991,Shu:1994,Hartmann:1994,Hayashi:1996}. 
Indeed, \citet{Favata:2005} speculated that the large magnetic loops 
observed in the {\sc coup} sample may be facilitating this type of magnetic 
star-disk interaction, in part because they argued that such large 
loops would likely be unable to remain stable if anchored only to 
the stellar surface.  However, as most of the 32 {\sc coup} sources 
studied by \citet{Favata:2005} lacked at that time sufficient photometric data 
with which to characterize the optical-infrared ({\sc ir}) spectral 
energy distributions, they could not confirm the presence of inner disks to which the 
observed magnetic loops might link.

Thus, there is still an outstanding question as to whether stellar 
coronal activity in these stars alone can drive such energetic flare 
events, or whether the energy (or at least the trigger) derives from 
a star-disk interaction.  While the latter requires magnetic loops 
large enough to reach the inner edge of the disk, several theoretical 
studies \citep[e.g.,][and references therein]{Uzdensky:2002,Matt:2005a}
have shown that the presence of a disk truncates 
the stellar magnetosphere so that closed magnetic loops extend not 
much further than the inner edge of the disk.  Thus, if these flares 
are powered by the star-disk interaction, one would expect the size 
of the flaring loops to approximately coincide with the location of 
the disk inner edge.  On the other hand, if the energetic flares are 
purely a stellar phenomenon, the largest loop sizes may only be 
exhibited by stars that lack disk material close to the star.  To 
address this question, it will therefore be useful to be able to 
determine the proximity of disk material to the furthest extent of 
the flaring magnetic loops exhibited by the sample analyzed in this work.

Near- and mid-{\sc ir} colors can be used as a crude tracer of close-in
circumstellar material. \citet{Getman:2008b} have used {\it Spitzer}
{\sc ir} colors to distinguish Class~{\sc ii} and Class~{\sc iii} objects 
(i.e., stars with dusty disks and naked T~Tauri stars, respectively) 
among 161 flaring {\sc coup} stars. Interestingly, they found evidence that
whereas the largest flaring loops tended to be associated with Class III 
objects, the Class~{\sc ii} sources in their sample were more likely to possess 
relatively small magnetic loops. The {\sc ir} colors alone do not provide
a quantitative measure of the location of the inner disk edge, but
\citet{Getman:2008b} suggest that the magnetic loops may be confined
by the disk to be within the disk co-rotation radius (the radius at
which disk material, if present, orbits the star with angular velocity
equal to the star's angular velocity). This suggestion is important,
as the disk co-rotation radius is the point specifically at which
some magnetospheric accretion theories predict magnetic star-disk
interaction to occur \citep[e.g.,][]{Ostriker:1995}. It is desirable,
therefore, to establish the relationship between circumstellar disks
and the very large magnetic loops observed by {\sc coup} more quantitatively 
than {\sc ir} colors alone permit.

In this paper, we present detailed spectral energy distributions ({\sc sed}s) 
for each of the 32 most powerful {\sc x}-ray flaring {\sc coup} sources 
at wavelengths 0.34--8~$\mu$m, plus upper limits at 24~$\mu$m 
(\S\ref{data}). In \S\ref{excessive} we present near-infrared color excesses
for the sample as a basic tracer of close-in circumstellar disks.
Next we compare in detail the full observed {\sc sed}s against synthetic {\sc sed}s
of low-mass {\sc pms} stars with disks (\S\S\ref{models}--\ref{example})
in order to (a) ascertain whether dusty disks are present 
around these stars, and (b) if so, determine quantitatively whether the 
inner edges of those disks are sufficiently close to the stellar surface 
to interact with the large flaring loops observed by \citet{Favata:2005}.
The results (\S \ref{results}) indicate that approximately more than half of the sample 
stars lack significant disk material within reach of their flaring magnetic 
loops; evidently the extremely large flaring loops observed by \citet{Favata:2005} 
are in most cases free-standing structures anchored only to the stellar 
surface. In \S\ref{discussion} we discuss some implications of this finding.

\section{DATA\label{data}}

\subsection{Study Sample, Loop Heights, and Stellar Data\label{data2p1}}

The 32 stars for our study constitute a unique subset of the {\sc coup}
\citep{Getman:2005} observation, identified by \citet{Favata:2005}
as exhibiting the brightest $\sim 1$\% of all flares observed by {\sc coup}.
These 32 flares had sufficient photon statistics with which a uniform
cooling loop ({\sc ucl}) analysis could be performed.

The {\sc ucl} model is based on
observations of solar flares. The occurrence of reconnection events
on the Sun has been used to benchmark relationships between {\sc x}-ray
flare decay slopes and the magnetic field structure confining the
emitting plasma. After a magnetic reconnection event occurs, heated
plasma evaporates from the chromosphere into the confining loop. The
material then emits soft {\sc x}-rays as it cools \citep{Priest:2002},
and the {\sc x}-ray light curve's decay time as well as its slope in density-temperature
space is related to the magnetic loop length \citep{Reale:1997}. This
method was developed using hydrodynamic simulations which were calibrated 
against spatially resolved imaging observations of solar flaring loops.

\citet{Favata:2005} applied the {\sc ucl} analysis to their sample of 32
stars and thus derived the lengths of the magnetic loops confining the
flare events observed by {\sc coup}. 
A reanalysis by \citet{Getman:2008a} includes these 32 objects.
The derived loop lengths in both
studies are consistent within uncertainties, so we adopt the former
for consistency throughout.  For simplicity, we estimate the loop height 
from the stellar surface as the loop length divided by 2.  This is an upper 
limit; for example, in a circular or semi-circular loop geometry, the actual 
loop height would be the loop length divided by $\pi$.  
These loop heights (half loop lengths) and their uncertainties are 
summarized in Table~\ref{stelparms}.
Uncertainties in the loop lengths are due to uncertainty in the 
measurement of the flare's peak temperature, decay time, and decay slope.  
For a detailed discussion of the quoted uncertainties in the loop lengths, see 
\citet[][cf.\ their $\S$3.2]{Favata:2005}.  

To narrow the range of acceptable best-fit spectral energy distributions (\S\ref{models}),
we require basic stellar parameters including 
effective temperatures ($T_{\rm eff}$), and radii ($R_{\rm star}$). 
These are taken primarily from \citet{LAH:1997} and are summarized
in Table~\ref{stelparms}. In cases where stellar parameters were not available 
from the literature, we adopt the temperatures and radii of the best-fit {\sc sed} model 
(see $\S$\ref{models}). Table~\ref{stelparms} 
also contains \ion{Ca}{2} equivalent widths \citep[as measured by][]{LAH:1997} 
and our newly reported \dik\ and \duv\ excess measurements
(see \S\ref{excessive}). We use \dik\ excess as a supplemental indicator 
of close-in, hot disk material, and the \ion{Ca}{2} equivalent widths in combination 
with \duv\ excess to indicate ongoing accretion onto the stellar surface.

\subsection{Photometric Data \label{phot}}

Fluxes for each of the 32 stars in our study sample were assembled over
the wavelength range 0.34~$\mu$m ({\it{\sc u}} band) to 24~$\mu$m ({\it Spitzer}
Multiband Imaging Photometer, {\sc mips}). With these data, we probe the 
stellar photosphere and circumstellar dust content. These measurements are 
summarized in Tables~\ref{fluxes1} and \ref{fluxes2}.

Optical fluxes were taken from the ground-based observations of the
Orion Nebula Cluster ({\sc onc}) by \citet{DaRio:2009} in the $UBVI_C$
passbands (0.36, 0.44, 0.55, and 0.83~$\mu$m, respectively), obtained
with the {\sc eso} Wide Field Imager ({\sc wfi}). We supplemented these with 
fluxes from {\it Hubble} Space Telescope Advanced Camera for Surveys ({\sc acs}) 
data \citep{Robberto:2005}, providing broadband fluxes at 0.43, 0.54, 0.77, and 
0.91~$\mu$m, as well as $V$ and $I_C$ magnitudes from 
the ground-based observations of \citet{LAH:1997}.

The 2 Micron All Sky Survey \citep[2{\sc mass};][]{Skrutskie:2006} provides
near-infrared $JHK_S$ magnitudes. Of critical importance to our analysis,
infrared photometry from the \textit{Spitzer} Infrared Array Camera
({\sc irac}) and {\sc mips} instruments provide the clearest probes of warm
circumstellar dust in the 3.6, 4.5, 5.8 and 8~$\mu$m bandpasses ({\sc irac})
and at 24~$\mu$m ({\sc mips}). We measured these
fluxes using pipeline-processed, archival data, and found our values to
agree within a few mJy of the unpublished measurements of the Spitzer {\sc gto} team 
(S.~T.\ Megeath, private communication). The {\sc mips} fluxes were measured by us
from the {\it Spitzer} archive using the pipeline reduced 24~$\mu$m images.
Unfortunately, the {\sc mips} image of the {\sc onc} is saturated over most of
the region of interest, and we were thus unable to recover more than a
few upper limits (see Table~\ref{fluxes2}).

Where magnitudes were originally reported, these have been converted
to fluxes using published zero points for each instrument (see final
rows in Table~\ref{fluxes1} and \ref{fluxes2}). In addition, we have in
general adopted larger uncertainties on the fluxes than the
formal measurement errors, in order to account for typical variability
levels in the optical and near-infrared of $\sim 0.1$ mag 
\citep[e.g.,][]{Herbst:1994,Carpenter:2001}. Specifically, we adopt an uncertainty 
of at least 10\% on the fluxes, unless the formal measurement error is larger.

\section{PRELIMINARY DISK DIAGNOSTICS: COLOR EXCESS AND ACCRETION INDICATORS
\label{excessive}}

Traditionally, the presence of warm circumstellar dust around
low-mass {\sc pms} stars has been traced using near-{\sc ir} ``color excesses,''
such as $\Delta(H-K)$, defined as the difference between the observed
(de-reddened) color and the color expected from a bare stellar photosphere
\citep[e.g.,][]{Strom:1989,Lada:1992,Edwards:1993,Meyer:1997}. The use of a 
single color excess of course does not permit a detailed, quantitative determination 
of disk structure (such as the size of the inner truncation radius, which is 
our primary interest here) because a given color excess depends in complex
ways upon multiple disk and stellar parameters (see also \S\ref{models} below).
Moreover, near-{\sc ir} colors may cause the observer to miss the presence of some disks, 
particularly those with large inner holes or around very cool stars whose 
photospheres peak in the near-{\sc ir}. For example, the overall disk frequency in 
the Orion Nebula Cluster has been estimated at $\sim$65\% on the basis of excess 
emission in the K band (Hillenbrand et al. 1998), but increases to $\sim$85\% simply 
by adding an L-band measurement (Lada et al. 2000). In other words, the addition 
of mid-{\sc ir} measurements can be very important for the detection of disks
\citep[for a discussion of disk-detection efficiency using near-{\sc ir} colors,
see][]{LAH:1998,Lada:2000,Ercolano:2009}.  
This is particularly relevant to our sample, in which many of the stars have 
cool photospheric temperatures, and where even disks with relatively large
inner holes could be within reach of the observed very large magnetic flaring 
loops (see Table~\ref{stelparms}).

Still, color excesses have the 
advantage of being easy to collect and analyze for large numbers of 
stars---especially prior to the advent of the wide and deep longer-wavelength
surveys made possible by {\it Spitzer}---and of providing a relatively
straightforward ``yes/no'' criterion for the presence of a disk.
For example, \citet{LAH:1998} used the $\Delta(I-K)$ color excess to conduct 
a census of disks among $\sim 1000$ low-mass stars in the {\sc onc}.
In that approach, the observed $V-I$ color was used to measure the
extinction, $A_V$, which was used in turn to deredden the observed $I-K$
color. Any excess $\Delta(I-K)$ was then attributed to the presence of a disk.
This approach has the advantage of requiring for each star only three flux 
measurements ($VIK$) and a spectral type (with which to establish the expected
photospheric colors). It assumes that (a) the observed $I$-band flux is purely
photospheric in origin, and (b) the observed $V$-band flux is only affected by 
reddening (i.e.,\ does not include any ``blue excess'' due to veiling emission
from accretion).

Using our compiled fluxes in Tables~\ref{fluxes1} and \ref{fluxes2} for
our study sample, we have calculated \dik\ in a manner
similar to \citet{LAH:1998}. However, rather than use only the observed 
$V-I$ color to determine $A_V$, and rather than normalize the stellar flux to 
the observed $I$-band flux, we have performed a two-parameter fit to each 
star's observed {\sc sed}.  To isolate the stellar flux from the disk and/or 
accretion flux, we use fluxes which appear to be photospheric in origin 
only, excluding the bluest wavelength fluxes which could be affected by accretion 
flux or scattered light \citep{Whitney:b2003} as well as the reddest wavelength 
fluxes which could contain flux from a disk. In general we used the fluxes from 
0.5 to 1.0~$\mu$m (total of 8 flux measurements, see Tables \ref{fluxes1} and 
\ref{fluxes2}) for this fitting, which should be a substantial improvement over 
the two-band ($V$ and $I_C$) approach described above.  To the observed fluxes we 
fit a {\sc NextGen} model atmosphere \citep{Hauschildt:1999} at the 
spectroscopically determined $T_{\rm eff}$ from the literature (Table~\ref{stelparms}). 
The two free parameters of the fit are the $A_V$ and the overall normalization of 
the stellar flux.  For comparison, previously published $A_V$ values \citep{LAH:1997} 
are listed in Table~\ref{stelparms}, and our newly determined $A_V$ values 
are also reported along with their corresponding errors (99\% confidence limits)
as determined from the two-parameter {\sc sed} fit. For 10 stars in our study sample,
$A_V$ values are reported here for the first time.
In many of these cases we find large $A_V$ values ($A_V \gtrsim 10$). 
These stars were likely absent from the optical study of \citet{LAH:1997} due
to the high reddening/extinction.

\begin{figure}[ht]
\includegraphics[width=6.0in]{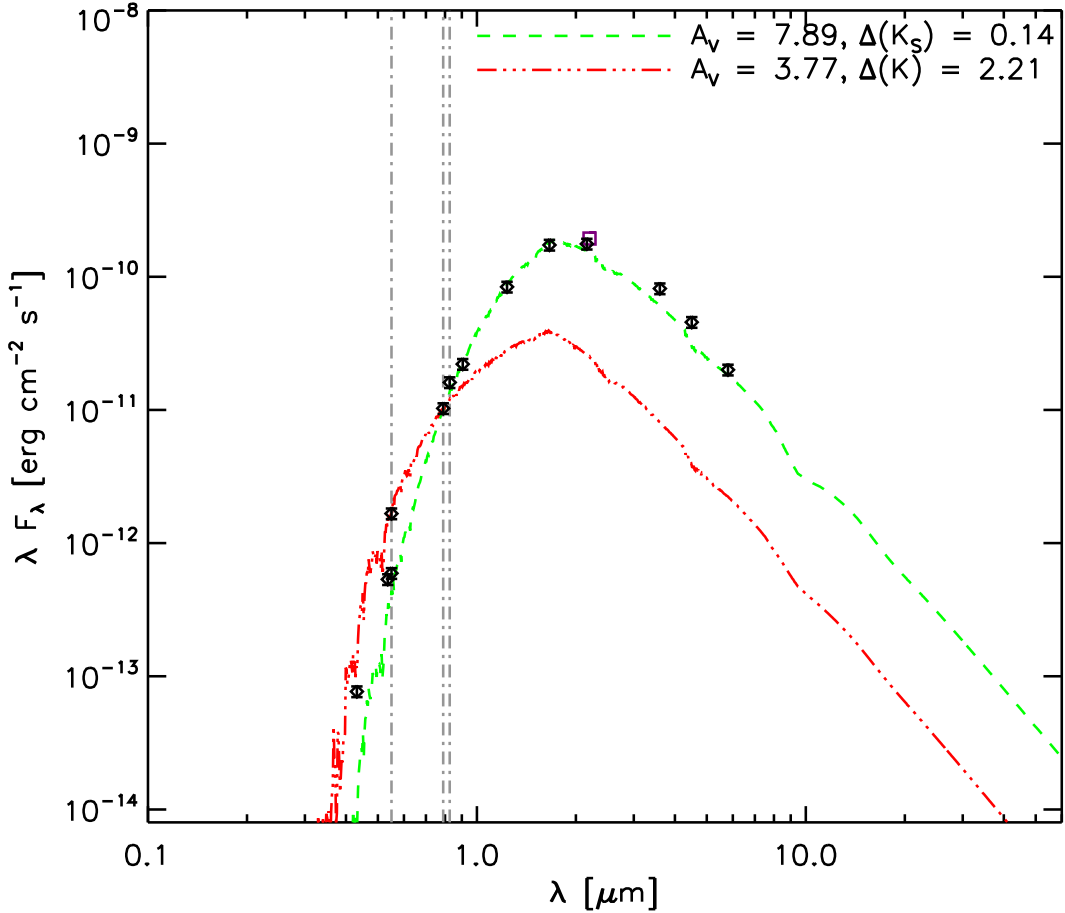}
\caption{\label{fig_262} 
{\sc sed} for {\sc coup}~262.  Diamonds are photometric data from 
Tables~\ref{fluxes1}--\ref{fluxes2} (see \S \ref{phot}). 
The purple square is the $K$-band flux from \citep{LAH:1998} 
used to calculate \dik\ for the red atmosphere model.
The green curve is the best fitting {\sc NextGen} stellar atmosphere model
with $T_{\rm eff}$ set to the literature value (Table~\ref{stelparms}).
The best-fit extinction, $A_V$, is reported upper right.  
The red curve represents the same stellar atmosphere model but with 
$A_V$ as previously determined by \citet{LAH:1997} based on a fit to the
$V$ and $I$ fluxes only. Vertical lines indicate the wavelengths of 
the (from left to right) $V$ and $I_C$ bands used by \citet{LAH:1997} 
and the {\sc wfi} $I$-band newly reported here.
The resulting \dik\ color excesses for both model atmosphere fits
are reported at upper right (see also Table~\ref{stelparms}).
It is clear that the new fit to the full set of available photometric fluxes
results in a more accurate representation of the stellar {\sc sed}. Whereas
this star was previously identified as possessing a very large near-{\sc ir}
excess, the new {\sc sed} fit here indicates no significant excess.
}
\end{figure}

For about half of the stars in our sample, our newly measured $A_V$ values 
agree within 99\% confidence with those previously reported. For the remaining 
stars, the $A_V$ values differ significantly; the reason for this difference 
is illustrated in Fig.~\ref{fig_262} for the case of star {\sc coup} 262.
The red model {\sc sed} represents the previously reported fit of a 
$T_{\rm eff} = 4395$~K photosphere to just the $V$ and $I_C$ band fluxes
from \citet{LAH:1997}.  The extinction that results is $A_V = 3.77$
\citep[see Table~\ref{stelparms} and][]{LAH:1997}, and as a consequence
the $K$-band flux appears to be highly in excess of the photosphere with
$\Delta(I-K) = 2.21$ \citep{LAH:1998}. However it is evident from visual inspection 
of the complete set of observed fluxes that this model fit is a poor representation 
of the additional measurements included here. As shown in the figure, our new fit 
to the entire set of fluxes gives $A_V = 7.91^{+0.70}_{-0.53}$ and \dik\ $= 0.14 \pm 0.11$.
For this particular case, evidently the previously reported $V$-band flux was 
anomalously high by $\sim 5\sigma$ (perhaps due to the ubiquitous optical 
variability of {\sc pms} stars), and so fitting for $A_V$ to just the $V$ and $I_C$ fluxes resulted
in a distorted model {\sc sed}. The difference between the two {\sc sed} fits
is important in the context of our study: e.g., the previously reported value of 
$\Delta (I-K)$ for {\sc coup} 262 implies a massive, warm circumstellar disk close to 
the star, whereas our newly determined best-fit {\sc sed} model is in fact consistent 
with no close-in disk. We revisit the {\sc sed} fit of {\sc coup} 262 in the context 
of the entire study sample below (\S\S~\ref{models} and \ref{results}).

\begin{figure}[ht]
\includegraphics[width=6.5in]{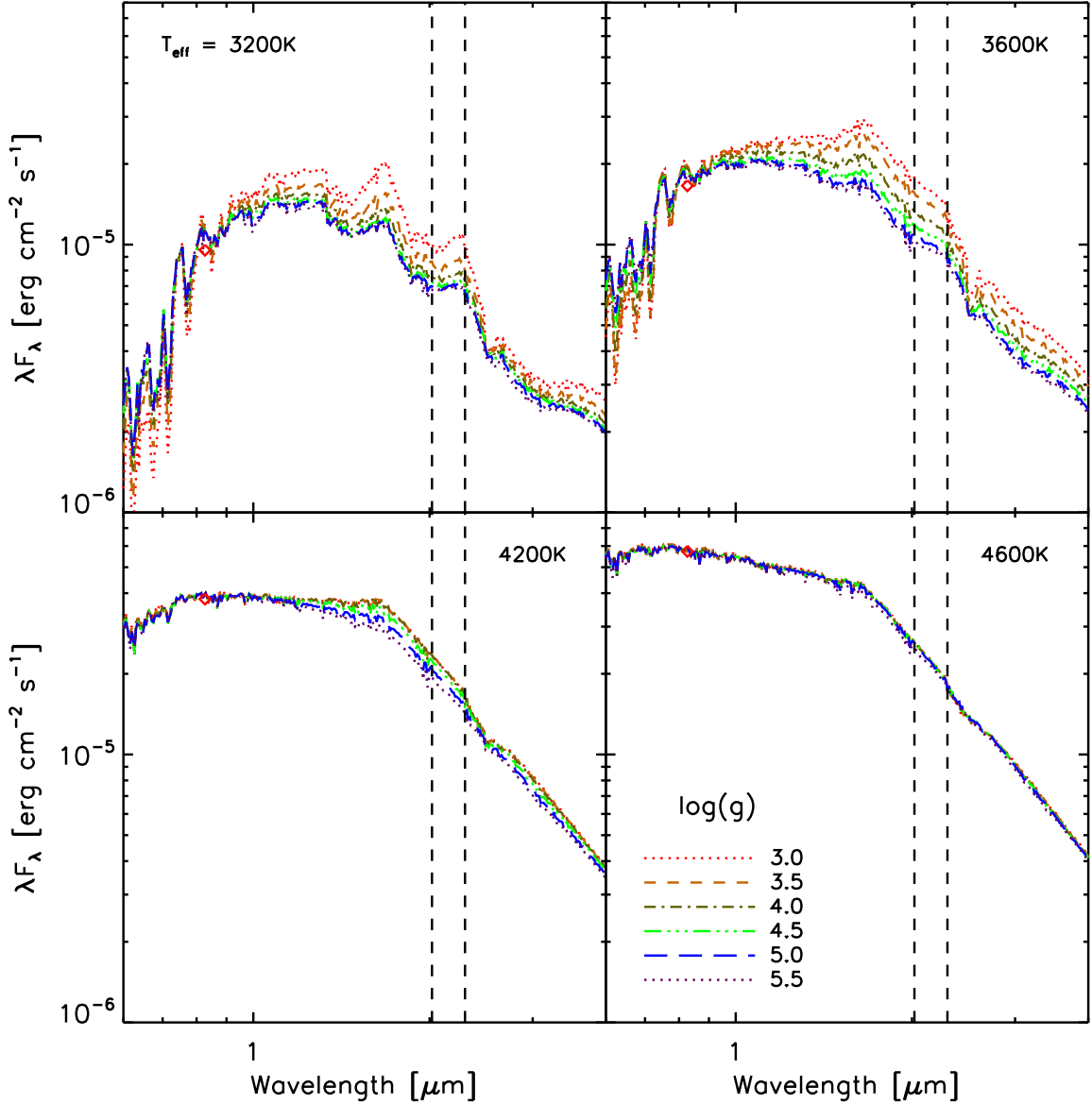}
\caption{\label{fig_loggs} 
Effects of changing surface gravity ($\log g$) on flux in the $K_S$ 
passband (represented by vertical dashed lines) as a function of
stellar temperature.  Each pane represents a different temperature
within the expected range for young, low-mass stars in our study
sample. In each plot, solar-metallicity {\sc NextGen} stellar 
atmospheres are plotted with six different $\log g$ values. Each 
atmosphere is normalized to the $I_{\rm WFI}$ bandpass at 0.83~$\mu$m,
indicated by red diamonds.  For cooler stars, i.e.,\ the atmospheres 
in the upper panels, $K_S$ flux varies by a factor of three depending 
on $\log g$; the effect is most pronounced at 
$T_{\rm eff} \lesssim 4000$ K.
}
\end{figure}

\citet{LAH:1997} adopted photospheric colors of main-sequence dwarfs 
(i.e.,\ for dwarfs, $\log g \sim 4.5$) in the calculation of near-{\sc ir} excesses. 
However, low-mass {\sc pms} stars at the young age of the {\sc onc} ($\sim 1$~Myr) are 
expected to have somewhat lower $\log g$ valuesdue to their large radii. Thus
we have also considered the extent to which the assumed $\log g$ affects 
the predicted stellar colors and thus the inferred near-{\sc ir} excesses. 
In Fig.~\ref{fig_loggs}, we compare the {\sc sed}s of
{\sc NextGen} stellar atmosphere models as a function of $\log g$ for
several representative $T_{\rm eff}$ appropriate for our study sample (all 
models shown are solar metallicity).
We see that the choice of $\log g$ is not important for stars warmer than
$T_{\rm eff} \gtrsim 4200$~K. However, we find that for cooler objects the 
predicted $I-K$ colors become increasingly redder with decreasing $\log g$, which
could lead to over-estimated near-{\sc ir} excesses for the coolest stars. 
Therefore
in this study we have opted to use model atmospheres with $\log g$ appropriate
to each object ($\log g$ calculated from previously reported stellar masses and 
radii; see Table~\ref{stelparms}).

Newly determined \dik\ values following the procedure described above
are reported in Table~\ref{stelparms}.  Several previous studies
\citep[e.g.,][]{Stassun:1999,Rebull:2001,Herbst:2002,Lamm:2004,Makidon:2004},
adopted a threshold value of \dik\ $> 0.3$ (i.e.,\ $\sim 3\sigma$
excess given typical $\sigma_K = 0.1$ mag; see Sec.~\ref{phot}) for
identifying stars with close-in circumstellar disks.  By this
criterion alone, five of the stars in our sample (COUP~141, 223, 1246,
1343, 1608) show large near-{\sc ir} excesses indicative of the presence of
warm circumstellar dust
\footnote{As described below, the large \dik\ in COUP 1246 is likely
the result of an underestimated $A_V$, and we consider its near-{\sc ir}
excess to be not significant for the remainder of our analyses.}.
The remaining stars in our sample show very weak or no evidence for
near-{\sc ir} excess emission (i.e.,\ \dik\ $< 0.3$).

In the context of the principal aims of the present study---where we seek
to determine whether the large magnetic loops observed in
the sample stars are linked to circumstellar disks---one might anticipate that 
\dik\ could be used as a quantitative tracer of circumstellar 
dust located within reach of the observed magnetic loops. For example, if 
the presence of substantial \dik\ excess correlates with the location of the 
disk truncation radius ($R_{\rm trunc}$, the distance from the inner edge of the 
disk to the star), then we might simply take stars with \dik\ $> 0.3$
as those whose inner disks are likely to be magnetically linked to the star.
This is similar to the approach of \citet{Getman:2008b}.

The combination of factors discussed above, however---relatively cool
stellar photospheres which peak in the near-{\sc ir}, magnetic flaring loops
that are large enough to interact with disks at relatively large
heights above the stellar surface, etc.---makes near-{\sc ir} excess an
inefficient tracer of the types of disks we seek to characterize. 
Table~\ref{foursquare} categorizes the sample stars
according to whether or not they display significant near-{\sc ir} excess emission
(i.e.,\ \dik\ $> 0.3$ vs.\ \dik\ $< 0.3$) and whether the inner-disk
truncation radius is larger or smaller than the dust destruction radius
(i.e.,\ $R_{\rm trunc} > R_{\rm dust}$ vs.\ $R_{\rm trunc} \lesssim R_{\rm dust}$, 
where $R_{\rm dust}$ is the distance from the star within which dust is warm 
enough to sublimate). Here $R_{\rm trunc}$ is determined from our detailed 
{\sc sed} model fitting as described below (\S\ref{models}).
The off-diagonal elements of Table~\ref{foursquare} represent cases contradicting
the assumption that \dik\ excess correctly and quantitatively predicts the 
location of the inner-disk edge.  
For only one of the sample stars (COUP 1246) do we find a relatively large
inner-disk hole ($R_{\rm trunc} > R_{\rm dust}$) but a large \dik\ $> 0.3$.
This one case appears unusual because the best-known explanation for a
strong near-{\sc ir} excess is the presence of warm dust close to the star.
Upon closer inspection of this case (COUP 1246), we found that the fitted $A_V$ 
was underestimated because of the presence of a moderate blue excess, likely 
due to chromospheric activity as suggested by the observed filled-in \ion{Ca}{2} 
emission (Table \ref{stelparms}), and which is not included in the photosphere 
model. If we manually adjust the $A_V$ value upward by $\sim3 \sigma$ from the fit
value of $A_V=1.52^{+0.70}_{-0.60}$ (see Table \ref{stelparms}) to $A_V=2.22$, 
the value of \dik\ becomes 0.17.  Indeed, our final best fit {\sc sed} model (see
\S\ref{models}) has an $A_V$ of 3.04, and thus an even lower \dik. 
Therefore, for the following analysis and discussion, we assume that
the calculated excess \dik\ in COUP 1246 is not significant (i.e.,
\dik\ is consistent with being less than 0.3).

Several stars in our sample exhibit a meager \dik\ $< 0.3$, and by
this criterion alone would be classified as lacking close-in, hot dust. 
However, for eight of these stars (in the lower left quadrant of Table 4), our
{\sc sed} fitting found these to in fact have disks that reach relatively
close to the star ($R_{\rm trunc} \lesssim R_{\rm dust}$), and in some
cases (shown below, \S\ref{results}), the disk reaches sufficiently
close to the star to interact with the observed large flaring loops.
In summary, Table 4 indicates that while \dik\ $\ge 0.3$ appears to be
an accurate indicator of dusty material close to the star, the lack of
significant \dik\ does not rule out the presence of dusty material
close to the star.  Thus, for the purposes of the present study where
we seek to establish more quantitatively the location of the inner-disk
edge in relation to the observed large magnetic flaring loops, we
cannot rely solely on traditional near-{\sc ir} excesses.

In addition to \dik\ measurements, from our new {\sc sed} fits we calculate blueward 
color excesses, $\Delta (U-V)$ which can be used as a tracer of accretion at the 
stellar surface (``hot spots'') and/or chromospheric activity and can help in our 
interpretation of some {\sc sed}s (see \S \ref{example}; these are reported in 
Table~\ref{stelparms}.).  Table~\ref{foursquare2} categorizes the sample stars in a 
manner similar to Table~\ref{foursquare}, but now using \duv.  We identify objects with 
\duv\ $< -0.3$ as those likely possessing hot accretion spots on their surfaces 
\citep[see ][]{Rebull:2000} and thus likely to be undergoing active accretion 
\citep[but see ][for a discussion of other phenomena that may cause blue excesses in 
{\sc pms} stars]{Findeisen:2010}.
Five of the sample stars show evidence for active accretion, and all but
one of these have $R_{\rm trunc} \lesssim R_{\rm dust}$ as expected for a
disk that extends close enough to the star for accretion to occur. Furthermore
for two of these stars the \ion{Ca}{2} measurements of 
\citet[][see Table~\ref{stelparms}]{LAH:1997} also indicate active accretion.
Only one star (COUP 1568) fails to show \duv\ excess despite possessing
a close-in disk edge. Thus, while \duv\ cannot provide a quantitative measure
of the location of the inner disk for non-accretors [\duv\ $> -0.3$], it is as 
expected a relatively reliable indicator of $R_{\rm trunc} \lesssim R_{\rm dust}$
for active accretors [\duv\ $< -0.3$].

In the analysis that follows, we use detailed {\sc sed} fits over the full range 
of available photometric data (Tables~\ref{fluxes1} and \ref{fluxes2}).
Where applicable, we use the \dik\ and \duv\ excesses and \ion{Ca}{2} 
equivalent widths in Table~\ref{stelparms} 
as secondary information to aid  our classifications in order to characterize 
in detail the presence and structure of circumstellar disks in our sample.

\section{SYNTHETIC SPECTRAL ENERGY DISTRIBUTION MODELS\label{models}}

To compare the observed {\sc sed}s of our sample with the {\sc sed}s expected from young stars 
with disks within reach of the observed flaring loops, we employed the Monte Carlo 
radiative transfer code of \citet{Whitney:b2003,Whitney:a2003}, {\sc ttsre},
to generate synthetic {\sc sed}s. Our aim is to more quantitatively constrain
the structure of any circumstellar material around each of the 32
stars in our sample so that we may determine, in detail for each star,
whether there is in fact disk material within reach of the magnetic
loops observed by \citet{Favata:2005}. Thus for each of the stars in
our sample, we wish to determine the range of disk parameters---the
most important of these being the location of the inner edge of the
disk---that are able to reproduce the observed {\sc sed}s (\S\ref{data}).

The {\sc ttsre} code models randomly emitted photons from the central
illuminating source and follows the photons as they interact with
(i.e., are absorbed or scattered by) any circumstellar material. The
circumstellar material is modeled as an optically thick dust disk
extending from an inner truncation radius, $R_{\rm trunc}$, to an
outer radius of typically a few hundred {\sc au}. The disk in general may be
``flared'' such that its scale-height increases with increasing distance
from the star, or it may be flat. Surrounding the star and disk may be 
a spherically
distributed infalling envelope with bipolar cavities; such an envelope
is generally required for reproducing the scattered-light properties of embedded 
objects \citep[generally seen as moderate excesses in the blue; e.g.,][]{Stark:2006}. 
The code also self-consistently solves for thermal equilibrium in the disk
as absorbed photons heat the disk and are re-radiated. Sublimation
of dust is also included \citep[for details of the dust properties
used by the code, see Table 3 of][]{Whitney:a2003}. The code models
the central illuminating source using the {\sc NextGen} atmosphere
models of \citet{Hauschildt:1999}. We adopted the solar-metallicity 
atmosphere models, with $\log g$ and $T_{\rm eff}$ chosen according to
each star's observationally determined $M_{\rm star}$, $R_{\rm star}$, 
and $T_{\rm eff}$ (Table~\ref{stelparms}).

Because of the very large number of permutations on the possible
star/disk/envelope parameters included in the {\sc ttsre} model (i.e.,
disk mass, disk inner and outer radius, disk flaring profile, disk
accretion rate, disk inclination angle, etc.), there is in general
not a simple one-to-one correspondence between a given observed
{\sc sed} and, say, $R_{\rm trunc}$. Thus, to fully explore the range of
disk parameters that could possibly reproduce the observed {\sc sed}s
of our sample, we made use of the very large grid of {\sc ttsre}
models constructed by \citet{Robitaille:2006}. The grid includes
some 200,000 models representing 14 star/disk/envelope parameters
\citep[see Table~1 of][]{Robitaille:2006} that were independently
varied to encompass virtually all possible combinations for young
stellar objects with masses 0.1--50 M$_\odot$ in the Class~0--{\sc iii} 
stages of evolution. The parameter space for this grid was specifically set to 
be very finely sampled for T Tauri stars (i.e., $T_{\rm eff} < $5200 K); at 
higher temperatures, $\ga$5200 K, the grid is more sparsely sampled. Whereas 
at $T_{\rm eff} < $5200 K the grid is sampled in $R_{\rm trunc}$ by $\sim$5\%, at 
$T_{\rm eff} >$ 5200 K it is sampled much more sparsely at $\sim$50\%. 
This issue affected only the hottest object in our sample, {\sc coup} 597 
(see \S \ref{criteria}).  Additionally, by construction, the models in 
this grid are set to only include non-zero accretion rates and to always 
include emission from a hot accretion spot on the star. As a result, in 
some cases blueward excess due to the hot accretion spots is seen in the 
best-fit model {\sc sed}s, which we disregard when we lack $U$ or $B$ 
band fluxes to constrain the blue side of the {\sc sed}.

With the added free parameter of extinction, $A_V$, we 
searched the grid via $\chi^2$ minimization for all synthetic {\sc sed}s that fit the 
observed {\sc sed} of a given star within the 99\% confidence level \citep[that is, 
we rejected those models that yielded a $\Delta\chi^2$ goodness-of-fit likelihood 
of 1\% or less relative to the best-fit model;][]{Press:1995}. For stars with
spectral-type determinations from the literature (see Table~\ref{stelparms}),
we furthermore require the model fits to have stellar $T_{\rm eff}$ within
500~K of the literature value except for a few cases where we found it
necessary to relax the $T_{\rm eff}$ constraint in order to achieve an
acceptable {\sc sed} fit; these exceptions are noted when we discuss each object
individually below.

In addition to temperature, we also filter the best-fit models by disk
mass.  In fitting the {\sc sed}s of the sample stars, particularly in cases with
little or no {\sc ir} excess emission in the observed {\sc sed}, we found that a number
of the best-fit models nonetheless had disks with small R$_{\rm trunc}$, 
but only if the disk also had a very low mass. The extensive model grid of
\citet{Robitaille:2006} allows for disks with masses as low as $10^{-10}\;
{\rm M}_\odot$. Such low disk masses, however, may be well below what is physically 
realistic, and certainly below what is observable, for young T Tauri stars. Recent 
detailed studies of {\sc pms} stars with so-called ``transitional'' and ``pre-transitional''
circumstellar disks \citep[e.g.,][]{Espaillat:2007}---disks that
are undergoing the rapid disk-clearing process from the inside out
\citep[e.g.,][]{Barsony:2005}---show that even at this late stage
the circumstellar disks are in fact quite massive. For example,
\citet{Espaillat:2007} derive $M_{\rm disk} \approx 10^{-1}$ M$_\odot$ for
the pre-transitional disk of LkCa~15 with $R_{\rm trunc} \sim 45$ {\sc au},
and $M_{\rm disk} \approx 10^{-2}$ M$_\odot$ for the slightly more evolved
disk of UX~Tau~A, with $R_{\rm trunc} \sim 60$ {\sc au}. A more extreme case
is that of CoKu/Tau~4, for which \citet{DAlessio:2005} find an extremely
low $M_{\rm disk} \approx 10^{-3}$ M$_\odot$.  In what follows, we will
thus restrict our analysis to include only model {\sc sed}s with $M_{\rm disk}
> 10^{-3}$ M$_\odot$ as more accurately representing the empirical disks
of young, low-mass stars. For illustrative purposes, however, we display
all models with disk masses greater than 10$^{-4}$ M$_\odot$.

In addition, for the purpose of interpreting the resulting best-fit model 
{\sc sed}s, we found it useful for each star to generate an additional 
{\sc ttsre} synthetic {\sc sed} as a fiducial reference model. For the cases 
where there is apparent {\sc ir} excess in the data indicating the presence of a 
disk, we generate a fiducial model identical to the best-fit {\sc sed} model, 
except that we set $R_{\rm trunc}$ equal to the magnetic loop height, $R_{\rm loop}$ 
(Table \ref{stelparms}). In the cases for which the data show no {\sc ir} evidence 
for dusty disks, the fluxes of the best fit {\sc sed} model are essentially 
arbitrary beyond the longest wavelength data point.  For simplicity in these 
cases our fiducial model is a simple star$+$disk {\sc sed}, adopting stellar 
properties from the literature. The modeled structure is that of an optically 
thick, geometrically thin, slightly flared disk with no envelope and no 
accretion. The disk mass in these cases is set to 0.01 M$_{\Sun}$, and its inner 
truncation radius is again set equal to the magnetic loop height. Thus, in all 
cases, the fiducial model allows a direct, visual comparison of the observed 
and best-fit {\sc sed}s against that expected if the inner disk is within 
reach of the magnetic flaring loop. It is important to note that in cases 
where the magnetic loop height is within the dust destruction radius, the 
disk is truncated at dust destruction by default (i.e., the {\sc ttsre} models 
require $R_{\rm trunc} \ge R_{\rm dust}$).

We present in this work the {\sc sed}s 
of the 32 stars in our study sample. Flux measurements and upper limits
(Tables \ref{fluxes1}--\ref{fluxes2}) are represented by 
diamond symbols with error bars (which are in most cases smaller 
than the symbols) and blue triangles, respectively. The {\sc sed} of 
the underlying stellar photosphere ({\sc NextGen} atmosphere) is shown 
for comparison as a dashed line. Superposed on the observed {\sc sed}s,
the best-fitting {\sc sed} models from the 
grid of \citet{Robitaille:2006} discussed above are shown as dash-dotted curves.  
The fiducial {\sc ttsre} model that we calculated is shown as a solid, blue curve.
Figures\ \ref{fig_c1410}--\ref{fig_c997} are 
included in print for illustrative purposes below (see \S\ref{example}), with 
the remaining figures accessible electronically.

\section{INTERPRETING THE SPECTRAL ENERGY DISTRIBUTIONS\label{example}}

With observed and model {\sc sed}s in hand for all 32 stars in our study sample, 
we can now attempt to answer the central
question of this paper: Are the large flaring loops observed on these stars likely due 
to a magnetic star-disk interaction, or do they represent primarily stellar
phenomena unrelated to disks?  To answer this question, in this section we discuss 
the specific criteria by which we determine, from examination of each star's {\sc sed},
the likelihood that it possesses a disk whose inner edge is within reach of the 
observed flaring magnetic loop.

\subsection{{\sc sed} Categorization Criteria\label{criteria}}

As discussed above, in general we found that $R_{\rm trunc}$
correlates with $M_{\rm disk}$ in the model {\sc sed}s, e.g.,\ a 
small $R_{\rm trunc}$ can fit even a bare photosphere {\sc sed} if 
$M_{\rm disk}$ is made sufficiently small (see \S\ref{models}).
Thus we found it helpful to visualize $R_{\rm trunc}$ versus $M_{\rm disk}$,
as shown in the lower panel of Figs.\ \ref{fig_c1410}--\ref{fig_c997}
in order to better interpret
the {\sc sed} model fits.  A vertical line at 10$^{-3}$ M$_{\odot}$
indicates our disk mass threshold (see \S\ref{models}). In orange, we show
the stellar photosphere for reference.  In each figure, the $M_{\rm disk}$
and $R_{\rm disk}$ values corresponding to each model {\sc sed} from the upper
panel are shown as diamonds.  For comparison, the magnetic loop height
($R_{\rm loop}$) and its uncertainty (see \S\ref{data2p1}) are shown as 
a dashed line and a hatched region, respectively.

From these plots, we can thus begin to assess the degree of spatial
correspondence between $R_{\rm trunc}$ and $R_{\rm loop}$ for each
star. Cases for which $R_{\rm trunc} \le R_{\rm loop}$ can be interpreted
as representing disks that are within reach of the observed flaring
loops. Cases for which $R_{\rm trunc} > R_{\rm loop}$ are somewhat less
straightforward to interpret because we must first account for the effects
of dust sublimation. The location of dust destruction, or the sublimation 
radius, is calculated using each model's stellar temperature and radius 
as follows \citep{Robitaille:2006}:
\begin{equation}\label{rdust} 
R_{\rm dust} = R_{\rm star} \times \left(
\frac{1600~{\rm K}}{T_{\rm star}} \right)^{-2.1} 
\end{equation} 
where 1600~K is the dust sublimation temperature.  Uncertainties in $R_{\rm
star}$ and $T_{\rm star}$ create a range of possible values for $R_{\rm dust}$.
We adopt 5\% uncertainty in $T_{\rm star}$ and 20\% uncertainty in $R_{\rm star}$ 
in this calculation \citep[see e.g.,][]{LAH:1997}. In each of 
Figs.~\ref{fig_c1410}--\ref{fig_c997} (and in figures 3.1-3.32, accessible online), 
filled (red) diamonds represent models 
effectively truncated at the dust destruction radius, $R_{\rm dust}$, while 
unfilled points have vertical bars in the $-y$ direction to show where that 
particular model's dust destruction radius is located. For the model uncertainty 
in $R_{\rm trunc}$, we adopt 5\% and 50\% for stars with $T_{\rm eff} < $5200 K 
and $T_{\rm eff} > $5200 K, respectively (see \S \ref{models}). Thus, when we 
say the disk is truncated ``effectively'' at the dust destruction radius, the
intended meaning is that $R_{\rm trunc}$ and $R_{\rm dust}$ are equivalent within 
their uncertainties.  We also report fiducial $R_{\rm dust}$ values for our
sample stars in Table~\ref{stelparms}; these are calculated using
Eq.~\ref{rdust} and the $R_{\rm star}$ and $T_{\rm star}$ data in
that same table.

The broadband fluxes used here trace the spatial extent of a disk's dust; 
in principle the inner edge of the {\it gas} in the disk could extend even 
closer to the stellar surface.  For the cases in which we find that an observed 
dust disk is truncated at the dust destruction radius, the dust disk is likely 
truncated by sublimation, a process which would not remove gas.  Indeed, some 
systems have been observed to be accreting even though the dust disk is 
truncated far from the star \citep{Eisner:2005,Eisner:2007}.  For dust disks 
truncated near the sublimation radius, but not within reach of the magnetic 
loop, it is possible that a gas disk extends closer to the star and is 
truncated within the loop height 
\citep[this has been observed by][]{Najita:2003,Eisner:2005}.
Conversely, if the dust disk is truncated outside the dust destruction
radius, some other process may be responsible for clearing out the inner
portion of the disk, and therefore we assume that the inner gas is cleared
out as well \citep[e.g.,][]{Isella:2009}.

Finally, for a few stars we lack sufficient photometric data to adequately
constrain the location of $R_{\rm trunc}$. In most cases, this is due
to a lack of {\it Spitzer} photometry and thus the longest wavelength
measurement is the 2{\sc mass} 2.2~$\mu$m flux.  Consequently, the model {\sc sed}s
in these cases are largely unconstrained and result in a wide variety of
possible star-disk configurations which can fit the observed {\sc sed}.

Based on these considerations, in what follows we categorize our sample
stars into four groups, based on the degree to which the {\sc sed}s indicate
that the inner disk edge is within reach of the flare loop height:
\begin{description}
\item[Category 1:]$R_{\rm trunc} \le R_{\rm loop}$: The inner disk edge 
is clearly within reach of the magnetic flaring loop.
\item[Category 2:]$R_{\rm trunc} > R_{\rm loop}$ but 
$R_{\rm trunc} \approx R_{\rm dust}$:
The dusty inner disk edge is beyond the flaring loop height, however the
dust disk is truncated at the dust-destruction distance and thus
a gas disk may extend inward to $R_{\rm loop}$ 
(i.e., $R_{\rm trunc} \la R_{\rm dust}$).
\item[Category 3:]$R_{\rm trunc} > R_{\rm loop}$ and 
$R_{\rm trunc} > R_{\rm dust}$: The inner edge of the dust disk is clearly beyond
reach of the magnetic flaring loop.
\item[Category 4:]Indeterminate: More than one category above is permitted 
by the available data (generally due to lack of {\it Spitzer} data). 
\end{description}

Recall that we take half the loop length as the flare loop height, as a conservative upper limit 
(see \S\ref{data2p1}). In general, for a given star there are multiple {\sc sed} models that are 
good fits to the observed {\sc sed}, and in some cases the multiple best-fitting model 
{\sc sed}s yield a mixed verdict regarding the placement of $R_{\rm trunc}$ with respect 
to $R_{\rm loop}$. Thus if one of categories 1--3 above is favored by more than 
$\frac{2}{3}$ of the best-fit {\sc sed} models, we assign the star to that category,
and we assign ``indeterminate'' (category 4) otherwise.

\subsection{Example Cases\label{cases}}

\begin{figure}[ht]

\includegraphics[width=3.0in]{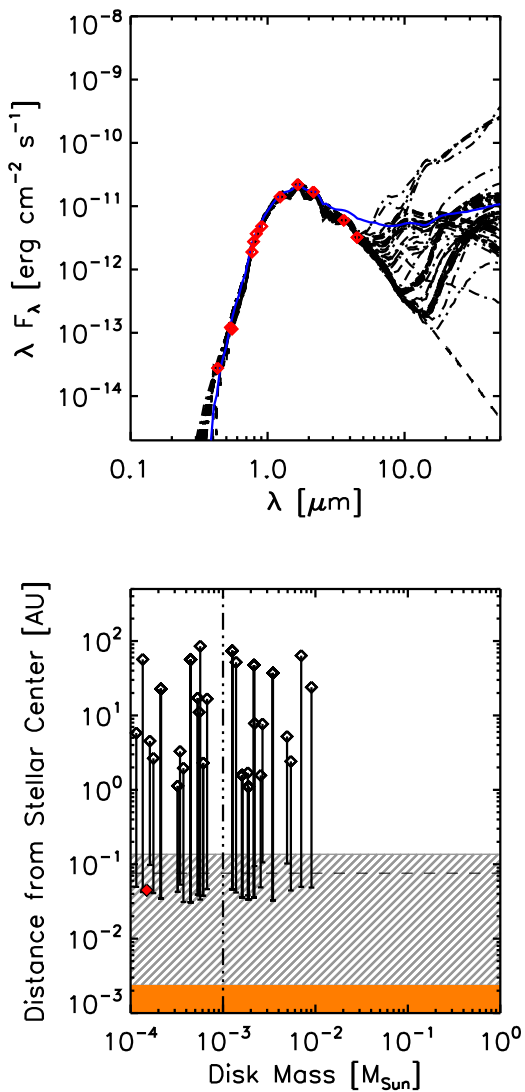}
\caption{\label{fig_c1410}
{\sc sed} of {\sc coup} 1410, category 3. All models are consistent with category 3; the SED is consistent 
with a bare photosphere.  This object is discussed in greater 
detail in $\S$\ref{cases}.  {\it Upper panel:} Best fit model {\sc sed}s from 
the model grid of \citet[][black dash-dotted curves]{Robitaille:2007}.  The 
fiducial {\sc ttsre} model with $R_{\rm trunc} = R_{\rm loop}$ (see \S\ref{models}), 
shown as a solid blue line and normalized to the peak near-{\sc ir} flux 
($J$, $H$, or $K_S$ band), is meant to illustrate approximately how the {\sc sed} 
would appear if there were a disk within reach of the flaring magnetic loop.  The 
dashed curve is a solar-metallicity {\sc NextGen} atmosphere model representing 
the stellar photosphere.  Red diamonds are measured fluxes 
(Tables \ref{fluxes1}--\ref{fluxes2}) as detailed in \S\ref{phot}.
{\it Lower panel:} Comparison of $R_{\rm loop}$ and $R_{\rm trunc}$ for
best-fit {\sc sed} models. A dashed line illustrates the loop height above the stellar surface 
(solid, orange).  Uncertainty in the loop height is 
shown as a gray, hatched region. Open diamonds represent the $R_{\rm trunc}$ values of 
the best-fit {\sc sed} models from the upper panel.  For each model, a vertical bar 
indicates the location of $R_{\rm dust}$ for that model (according to Eq.~\ref{rdust}).
Filled red diamonds indicate models which have $R_{\rm trunc} \approx R_{\rm dust}$
(see $\S$\ref{example} for more detail).  Finally, the vertical dash-dotted line 
indicates our disk mass threshold value, 10$^{-3}$ M$_{\odot}$; less massive disks do 
not represent the disks typical of T Tauri stars (see \S \ref{models}).  Figures for all 32
sample stars are available in color electronically (Figs. 3.1-3.32).}
\end{figure}

\begin{figure}[ht]

\includegraphics[width=3.0in]{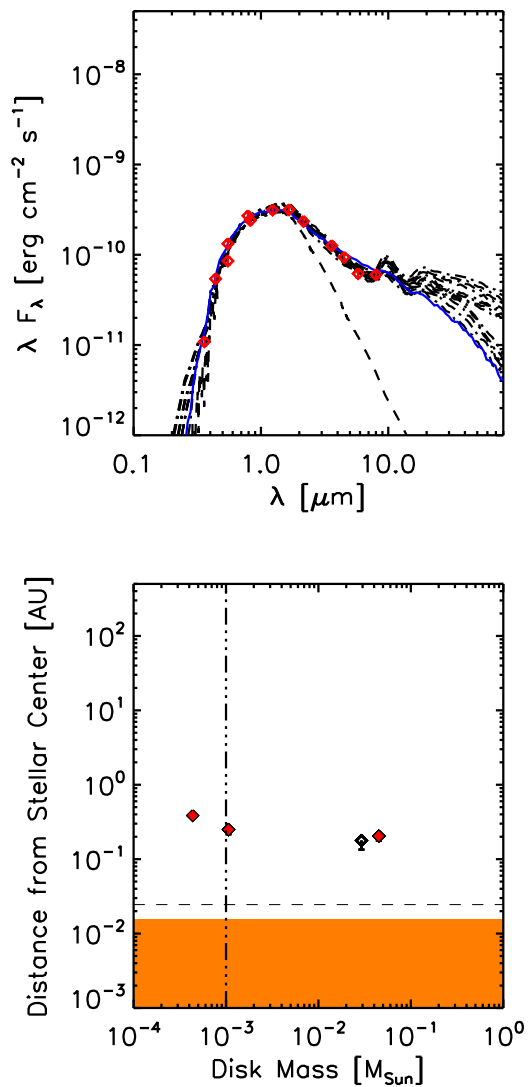}
\caption{\label{fig_c141}
{\sc sed} of {\sc coup} 141, category 2. All symbols are as in Fig.~\ref{fig_c1410}.
2/3 models are consistent with category 2.
Excess flux in the {\sc irac} bands indicates a dusty disk.  Fourteen best-fit 
{\sc sed}s are plotted, but three of these are degenerate in $M_{\rm disk}$ or
$R_{\rm trunc}$, representing four inclinations of the same star--disk configuration. 
Three sets of model {\sc sed}s have $M_{\rm disk} > 10^{-3}$ M$_\sun$, and more than
2/3 of these are truncated at their respective dust destruction radii (red points 
in lower panel). Thus these disks may possess gas that extends inward of 
$R_{\rm dust}$ to $R_{\rm loop}$.  This object is very likely accreting, based on 
its strong \ion{Ca}{2} emission ($-$17.8\AA) and $\Delta(U-V)$ excess of $-1.26$. 
For further discussion of this object, see $\S$\ref{cases}.}
\end{figure}

\begin{figure}[ht]

\includegraphics[width=3.0in]{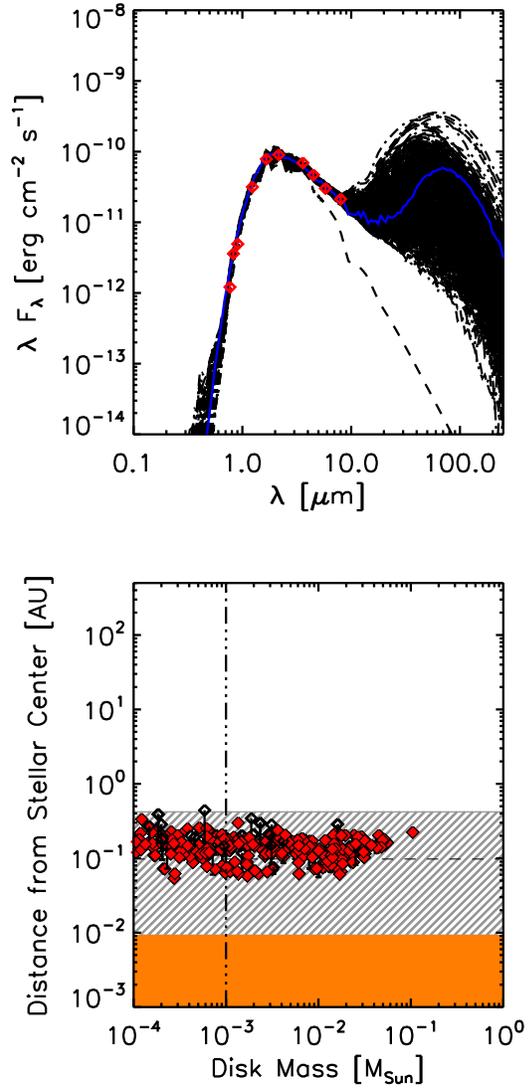}
\caption{\label{fig_c720}
{\sc sed} of {\sc coup} 720, category 1. All symbols are as in Fig.~\ref{fig_c1410}.
All 175 models are consistent with category 1.
This object is discussed in detail in \S\ref{cases}.}
\end{figure}

\begin{figure}[ht]

\includegraphics[width=3.0in]{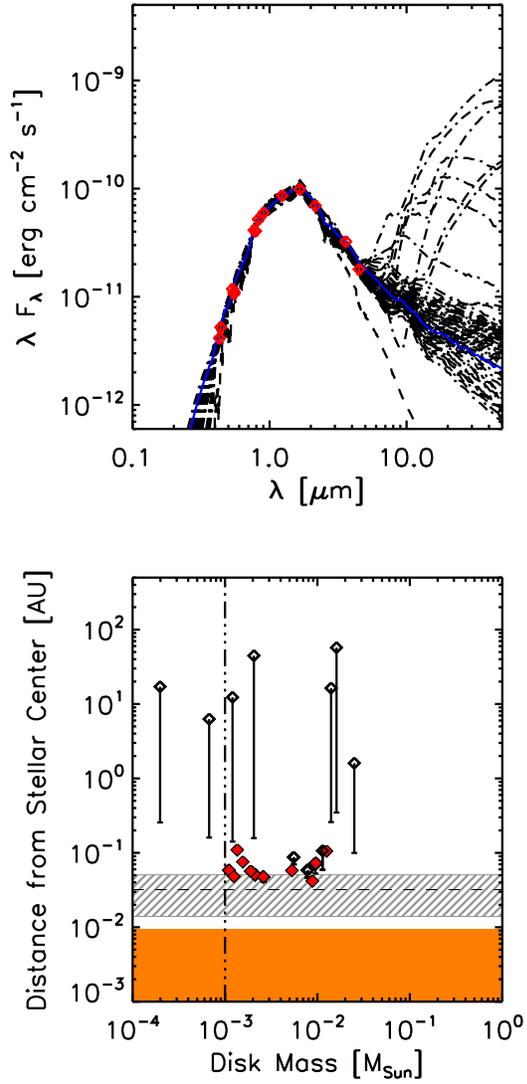}
\caption{\label{fig_c997}
{\sc sed} of {\sc coup} 997, category 4. All symbols are as in Fig.~\ref{fig_c1410}.
Most of the best fit model results are split between 
categories 2 (11/20) and 3 (9/20), placing this object into category 4. 
This object is discussed in detail in \S\ref{cases}.}

\end{figure}

As an example of our approach to interpreting the {\sc sed}s of our study
sample, we show in Fig.~\ref{fig_c1410} (upper panel) the {\sc sed}
of {\sc coup} 1410.  The fiducial {\sc sed} model in Fig.~\ref{fig_c1410},
corresponding to a disk with $R_{\rm trunc} = R_{\rm loop}$, predicts an
excess of {\sc ir} flux at wavelengths as short as 3~$\mu$m, unlike the data
and best-fit model {\sc sed}s which follow the profile of a bare stellar
photosphere to 4.5~$\mu$m.  Intuitively, this implies that the best-fit
model {\sc sed}s must therefore correspond to disks with moderately large inner 
holes.  Indeed, the lower panel of Fig.~\ref{fig_c1410} shows that nearly 
all of the best-fit model {\sc sed}s, representing disks with 
$10^{-4} \lesssim M_{\rm disk}/{\rm M}_\odot \lesssim 10^{-2}$, 
have $R_{\rm trunc} > 1$~{\sc au}.  Furthermore, the majority of
these models are truncated well outside their respective dust destruction
radii (i.e.,\ $R_{\rm trunc} > R_{\rm dust}$); only one fit model has 
$R_{\rm trunc} \lesssim R_{\rm dust}$, and this model has very low $M_{\rm disk}$, below our 
threshold of 10$^{-3}$ M$_{\odot}$

Note that the observed {\sc sed} for this star does not in fact {\it require} any disk
at all; the fact that many of the {\sc sed} models shown in Fig.~\ref{fig_c1410}
exhibit large excesses longward of 4.5~$\mu$m implies only that these
hypothetical disks with very large inner holes are formally {\it permitted} 
by the available data. These models thus provide a lower limit to the size
of $R_{\rm trunc}$ that any as-yet undetected disk could possibly have.
Since this lower limit is in this case much larger than $R_{\rm loop}$,
we conclude that no disk is present that could interact with the 
observed magnetic flaring loop, and we assign
{\sc coup}~1410 to category 3 ($\S$\ref{criteria}).

{\sc coup}~141 (Fig.~\ref{fig_c141}) is a case in which the observed {\sc sed} is 
reasonably well matched by the fiducial {\sc sed} model, for which 
$R_{\rm trunc} = R_{\rm loop}$.  The best-fit {\sc sed}s have inner truncation radii 
well beyond reach of the magnetic loop. However, these models' inner disk radii
are also equal to their dust destruction radii, and thus it is likely that
sublimation is responsible for the apparent clearing of the inner disk.  
In cases like {\sc coup}~141, while the magnetic loop may not intersect 
the dust disk, it could nonetheless intersect a gas disk that extends
inward of the dust to within reach of $R_{\rm loop}$. Indeed, both 
\ion{Ca}{2} and \duv\ strongly indicate active accretion (Table~\ref{stelparms}).
Thus we assign {\sc coup}~141 to category 2.

As another example, consider {\sc coup}~720 (Fig.~\ref{fig_c720}). In this
case, all of the best-fit model {\sc sed}s with $M_{\rm disk}$ above our adopted 
threshold of 10$^{-3}$ M$_\odot$ have similar $R_{\rm trunc} \sim 0.1$ {\sc au}, 
which overlaps $R_{\rm loop}$ within its uncertainty. Many of these best-fit models,
moreover, have $R_{\rm trunc} \approx R_{\rm dust}$, and thus
may possess gas disks that extend even closer to the star. {\sc coup} 720
thus represents a good example of an {\sc sed} that is consistent
with $R_{\rm trunc} \approx R_{\rm loop}$,
and for which the large magnetic loops observed by {\sc coup} may facilitate
the magnetic star-disk interaction envisaged in magnetospheric accretion
models. {\sc coup}~720 is assigned to category 1.

Finally, consider {\sc coup}~997 (Fig.~\ref{fig_c997}). The
observed {\sc sed} data (0.34--4.5~$\mu$m) show excess {\sc ir} flux. 
About half of the best-fit models are truncated at their dust
destruction radii (category 2), while the other half are truncated beyond 
1~{\sc au} (category 3). We also do not have \ion{Ca}{2} or \duv\ measurements
to help disambiguate the two possibilities, and thus
it is not possible to say which set of models correctly
describes the observed star-disk system.  
Requiring additional data (particularly longward of $\sim 10 \mu$m)
to discriminate between the category 2 and 3 model fits,
we assign this object to category 4.

\section{RESULTS\label{results}}

In Table~\ref{summary}, we present a summary of the results for the 32 stars in our 
sample. Following the procedure described in \S\ref{example}, we have identified which 
stars' {\sc sed}s have massive (i.e., $\geq$10$^{-3}$M$_{\odot}$) disks that are consistent 
with being within reach of the observed magnetic loops. Notes about each star relevant to 
its classification are provided in the figure captions.  

We find \catwone\ stars that appear to have {\sc sed}s consistent with
$R_{\rm trunc} \le R_{\rm loop}$ (category 1). Another \catwtwo\ stars
do not show direct evidence of disks within reach of the magnetic loops,
but could potentially have gas that extends interior to the observed dusty
inner edge of the disk (category 2).  \catwcthree\ stars either have disks
whose inner edges are situated beyond the reach of the magnetic loops, or
are simply devoid of detected disk material entirely (category 3). For \catwfour\
stars, we could not assign a definitive category as additional 
data are necessary to support or eliminate different classes of best-fit
{\sc sed} models (category 4).

\section{DISCUSSION AND CONCLUSIONS\label{discussion}}

Of the \goodguys\ stars in our 32-star sample for which we have enough 
optical--infrared data to constrain the location of the disk inner edge 
(i.e., excluding stars in category 4; \S\ref{criteria} and Table~\ref{summary}), 
for about \catpthree\ we are able to rule out close-in, massive disks within reach
of the observed large flaring loops (category 3). These energetic flares discovered 
by \citet{Favata:2005} are evidently intrinsically stellar phenomena. 
This gives added justification {\it a posteriori} for the application of 
the solar-flare cooling loop model to these stars (\S\ref{data2p1}), 
and suggests that it may be possible by further extension of the solar 
analogy to infer other flare-related properties for these flares, such as 
coronal heating rates and coronal mass ejections.  The latter
in particular may be important for furthering our understanding of
mass and angular momentum loss in these low-mass {\sc pms} stars.

Our sample also includes \catwone\ cases for which the {\sc sed} clearly indicates a
dusty disk that extends close enough to the star to permit interaction with the flaring 
loop (category 1; Table~\ref{summary}).  In \catwtwo\ additional 
cases the dust disk appears to be truncated beyond the reach of the 
flaring loop, but at or close to the predicted dust destruction radius 
(category 2).  In these cases, the dust disk may in fact be truncated by 
dust sublimation, a process which does not remove gas.  Thus, in category 2 objects, 
it seems likely that a gas disk (undetected in the broadband flux measurements used in 
our {\sc sed} models) extends closer to the star and may be within reach of the 
observed flaring loops.  One of the \catwtwo\ category 2 objects has 
a \ion{Ca}{2} measurement from the literature ({\sc coup} 141),
and two have $\Delta(U-V)$ measurements ({\sc coup} 141 and {\sc coup}
1568).  {\sc coup} 141 is, interestingly, the most strongly accreting
object in the sample as probed by its \ion{Ca}{2} equivalent width (see
Table~\ref{stelparms}) and it also has a very negative $\Delta(U-V)$;
the combination of these indicators is strong evidence for ongoing accretion. 
For {\sc coup} 141 the \dik\ near-{\sc ir} excesses also indicates a disk 
very close to the star (see Sec.~\ref{excessive} and Table~\ref{foursquare}). 
These examples further strengthen the interpretation of the category 2 stars as 
likely having gas accretion disks within reach of the stellar magnetosphere.

Several studies clearly find a high frequency of close-in, massive dusty 
disks in the {\sc onc} population as a whole. For example, \citet{LAH:1998} 
find a disk fraction in the {\sc onc} of $\sim$70\% on the basis of excess emission 
at 2.2~$\mu$m. Thus, the $\sim$\catpone\ (category 1) or $\sim$38\% (categories 
1 and 2) frequency of close-in disks in our 32-star sample, representing the 
$\sim$1\% of {\sc onc} stars with the most powerful X-ray flares observed by COUP, is 
evidently not representative of the disk characteristics of the {\sc onc} as a whole.

The \catwone\ cases in which we have observed stars with close-in, dusty disks 
which intersect the magnetic loops (category 1) are interesting candidates 
for further study.  Specifically, it would be informative to determine 
if the flares in these cases are in some way different than the category 3 
cases.  Three of these objects have \duv\ measurements, all of them $\le -0.3$ 
(Table~\ref{stelparms}), strongly indicating active accretion.  Temporally linking 
accretion as seen in optical variability to the X-ray flare events 
\citep[e.g.,][]{Stassun:2006}
could solidify whether a magnetic star-disk interaction has taken place.
Geometric information would be necessary to determine where the magnetic
loop is on the stellar surface---for example, a given flaring loop could
extend in a direction perpendicular to the disk and thus not interact,
even if the dust (or gas) disk is within the appropriate distance from
the stellar surface.

The question of how the large magnetic structures are stabilized (prior to
the flaring event) was posed in the discovery publication of
these objects \citep{Favata:2005}.  It was proposed that the loops may be
anchored to corotating disk material and thereby not subject
to shear which could disrupt the loops. In this work we have found 
such disk-anchoring to indeed be a possibility for 10 
objects (categories 1 and 2), but further analysis is needed
to determine how massive and ionized a disk must be to enable
interaction.  For \catwthree\ objects (category 3), disk-supported loops 
are unlikely, as these stars lack massive disks within reach of the loops.
We speculate that as long as the confining magnetic field at the upper end
of the loop is sufficiently strong and the confined material corotating,
stability is feasible even without a disk.   For example, \citet{Cranmer:2009} 
describes a loop geometry in which the pressure of the confined gas decreases 
with increasing loop length, implying that in fact the largest coronal loops 
may be most stable against rupture. Additionally, our findings may imply that 
the largest flaring loops cannot readily form in the presence of a disk,
given that they appear in our sample to preferentially occur on stars lacking 
close-in disk material.

Alternate scenarios to explain the long {\sc x}-ray decay
timescales include observations of corotating, embedded structures in coronae
\citep[e.g.,][]{Cameron:1989a,Cameron:1989b}, and prominences and stellar winds 
\citep{Skelly:2008,Massi:2008}.  The observed coronal structures from those 
studies lend additional support to the idea of these large magnetic structures 
remaining stable in hot coronae or within the stellar wind over multiple rotation 
periods as the {\sc x}-ray flare decays, even if no disk is present to anchor
the magnetic loop \citep{Jardine:2005}.

In summary, the 32 most powerful flares observed by the {\sc
coup} survey were found to have magnetic structures multiple stellar
radii in arc length confining the {\sc x}-ray emitting, heated plasma
\citep{Favata:2005}.
With the goal of understanding the nature of these large {\sc x}-ray
emitting flare structures, we have modeled the optical--infrared {\sc sed}s 
of these objects, finding \catpthree\ to be lacking close-in circumstellar 
disks to which these
loops could anchor.  It is evident that in at least these cases the
large-scale flares are phenomena of purely stellar origin, neither 
triggered nor stabilized by star-disk interactions.

\acknowledgments
This research is supported by NSF grant AST-0808072 (K. Stassun,
PI). K.~G.~S. gratefully acknowledges a Cottrell Scholar award from the
Research Corporation, and a Diversity Sabbatical Fellowship from the
Ford Foundation.  S.~P.~M. is supported by an appointment to the NASA
Postdoctoral Program at Ames Research Center, administered by Oak Ridge
Associated Universities through a contract with NASA. We thank S.~T.\ Megeath 
for kindly providing us {\sc irac} data for the {\sc onc} in advance of 
publication.

\clearpage

\begin{landscape}
\begin{deluxetable}{l|cccccccccc}
\tabletypesize{\scriptsize}
\tablecaption{Stellar Parameters\label{stelparms}}
\tablewidth{0pt}
\tablehead{\colhead{Object} & \colhead{T$_{\rm eff}$\tablenotemark{a}}  & \colhead{Mass\tablenotemark{a}}  & \colhead{Radius\tablenotemark{a}}	& \colhead{Magnetic Loop}                        & \colhead{Dust Destruction}                      & \colhead{\ion{Ca}{2} EW\tablenotemark{a}}   & \colhead{Literature}                      & \colhead{Re-calculated}                  & \colhead{$\Delta(K_{S})$\tablenotemark{c}} & \colhead{$\Delta (U-V)$\tablenotemark{c}}     \\
           \colhead{Name}   & \colhead{[K]}	                        & \colhead{[M$_{\sun}$]}           & \colhead{[R$_{\sun}$]}             & \colhead{Height [R$_{\sun}$]\tablenotemark{b}} & \colhead{Radius\tablenotemark{c} [R$_{\sun}$]}  & \colhead{[\AA]}                             & \colhead{A$_{V}$\tablenotemark{a} [mag]}  & \colhead{A$_{V}$\tablenotemark{c} [mag]} & \colhead{[mag]}                            & \colhead{[mag]}  }
\startdata
{\sc coup} 7    & 4581               & 2.12             & 6.23                    & 0.65                     &  56.8   &  \nodata   &  0.75      &  0.67 (-0.58, +0.54)  &     0.12	&   0.03      \\
{\sc coup} 28   & 3802               & 0.53             & 2.3                     & 7.91$\pm$5.46            &  14.2   &  1.6       &  0.63      &  0.30 (-0.30, +0.54)  &     0.20	&   0.24      \\
{\sc coup} 43   & 3606               & 0.4              & 2.92                    & 16.1 (+3.16,-5.32)       &  16.0   &  1.4       &  1.36      &  1.18 (-0.60, +0.56)  &     0.012	&   0.64      \\
{\sc coup} 90   & 3802               & 0.52             & 2.51                    & 1.04 (+4.86,-1.04)       &  15.5   &  1.6       &  4.97      &  3.97 (-1.13, +1.13)  &     0.053	&   \nodata   \\
{\sc coup} 141  & 5236               & 2.11             & 3.3                     & 1.97                     &  39.2   & -17.8      &  1.83      &  2.17 (-0.63, +0.60)  &     0.37	&  -0.67      \\
{\sc coup} 223  & 4395               & 1.19             & 2.79                    & 6.95                     &  23.3   &  1.7       &  4.66      &  5.81 (-0.75, +0.88)  &     0.31	&   \nodata   \\
{\sc coup} 262  & 4395               & 1.13             & 1.58                    & 28.5 (+16.8,-22.3)       &  13.2   &  2.3       &  3.77      &  7.89 (-1.20, +1.24)  &     0.14	&   \nodata   \\
{\sc coup} 332  & 3111$^{\dagger}$   & 0.5$^{\ddagger}$ & 2$^{\ddagger}$          & 105$\pm$38.4             &  8.09   &  \nodata   &  \nodata   &  12.8 (-1.67, +2.14)  &     0.032	&   \nodata   \\
{\sc coup} 342  & 4729$^{\dagger}$   & 0.5$^{\ddagger}$ & 2$^{\ddagger}$          & 20.3 (+21.4,-20.3)       &  19.5   &  \nodata   &  \nodata   &  7.89 (-0.78, +0.85)  &     0.085	&   \nodata   \\
{\sc coup} 454  & 4775               & 2.35             & 4.58                    & 46.4 (+14.8,-9.20)       &  45.5   &  2.1       &  5.85      &  6.39 (-0.85, +0.87)  &     0.077	&   \nodata   \\
{\sc coup} 597  & 5662               & 1.49             & 2.01                    & 3.16 (+7.62,-3.16)       &  28.6   &  4.5       &  2.69      &  3.32 (-1.35, +1.39)  &     0.27	&  -0.26      \\
{\sc coup} 649  & 3589               & 0.4              & 2.17                    & 9.20                     &  11.8   &  0         &  4.11      &  3.82 (-0.73, +0.83)  &     0.17	&   \nodata   \\
{\sc coup} 669  & 4581               & 1.52             & 2.59                    & 13.2 (+2.73,-3.31)       &  23.6   &  \nodata   &  1.96      &  2.33 (-1.10, +1.20)  &     0.049	&   \nodata   \\
{\sc coup} 720  & 4452$^{\dagger}$   & 0.5$^{\ddagger}$ & 2$^{\ddagger}$          & 19.0 (+69.3,-19.0)       &  17.2   &  \nodata   &  \nodata   &  11.8 (-1.33, +1.66)  &    -0.23	&   \nodata   \\
{\sc coup} 752  & 3802               & 0.54             & 1.67                    & 9.35 (+1.29,-9.35)       &  10.3   &  1.1       &  0.07      &  0.64 (-0.64, +1.10)  &     0.047	&  -0.70      \\
{\sc coup} 848  & 3342               & 0.29             & 1.98                    & 23.3 (+4.03,-3.31)       &  9.31   &  0         &  1.72      &  1.35 (-1.34, +1.46)  &     0.093	&   \nodata   \\
{\sc coup} 891  & 4775               & 2.43             & 4.85                    & 24.9 (+3.88,-3.31)       &  48.8   &  1.8       &  8.00      &  10.7 (-1.02, +1.12)  &     0.13	&   \nodata   \\
{\sc coup} 915  & 4613$^{\dagger}$   & 0.5$^{\ddagger}$ & 2$^{\ddagger}$          & 11.2 (+3.88,-2.59)       &  18.4   &  \nodata   &  \nodata   &  15.8 (-1.76, +2.56)  &    -0.065	&   \nodata   \\
{\sc coup} 960  & 3177               & 0.24             & 2.16                    & 0.53 (+1.48,-0.53)       &  9.27   &  0         &  2.72      &  1.29 (-0.41, +0.43)  &    -0.013	&   \nodata   \\
{\sc coup} 971  & 3999               & 0.69             & 3.28                    & 5.03 (+0.72,-0.43)       &  22.5   &  1.8       &  0.        &  0.00 (-0.00, +0.15)  &    -0.24	&  -0.80      \\
{\sc coup} 976  & 3177               & 0.18             & 0.91                    & 10.9                     &  3.85   &  0         &  0.        &  2.88 (-0.69, +1.13)  &     0.28	&   \nodata   \\
{\sc coup} 997  & 3856$^{\dagger}$   & 0.5$^{\ddagger}$ & 2$^{\ddagger}$          & 4.89 (+4.03,-3.88)       &  12.7   &  \nodata   &  \nodata   &  1.82 (-1.14, +1.13)  &     0.060	&   \nodata   \\
{\sc coup} 1040 & 4281$^{\dagger}$   & 0.5$^{\ddagger}$ & 2$^{\ddagger}$          & 1.25 (+1.91,-1.25)       &  15.8   &  \nodata   &  \nodata   &  16.5 (-2.03, +3.53)  &    -0.10	&   \nodata   \\
{\sc coup} 1083 & 4698$^{\dagger}$   & 0.5$^{\ddagger}$ & 2$^{\ddagger}$          & 33.8 (+9.63,-7.05)       &  19.2   &  \nodata   &  \nodata   &  4.16 (-0.56, +0.61)  &    -0.0072	&   \nodata   \\
{\sc coup} 1114 & 4903$^{\dagger}$   & 0.5$^{\ddagger}$ & 2$^{\ddagger}$          & 9.78 (+2.30,-9.78)       &  21.0   &  -1.5      &  \nodata   &  6.08 (-0.71, +0.74)  &     0.086	&   \nodata   \\
{\sc coup} 1246 & 3177               & 0.23             & 1.62                    & 5.75 (+1.01,-1.15)       &  6.95   &  0         &  0.92      &  1.52 (-0.60, +0.70)  &     0.45	&   \nodata   \\
{\sc coup} 1343 & 3649$^{\dagger}$   & 0.5$^{\ddagger}$ & 2$^{\ddagger}$          & 13.8 (+2.45,-2.44)       &  11.6   &  \nodata   &  \nodata   &  3.04 (-0.83, +3.66)  &     0.60	&   \nodata   \\
{\sc coup} 1384 & 3802               & 0.52             & 2.46                    & 7.33$\pm$4.46            &  15.1   &  1.9       &  0.        &  0.67 (-0.67, +0.82)  &    -0.095	&  -0.35      \\
{\sc coup} 1410 & 3606               & 0.36             & 0.51                    & 15.8 (+12.9,-15.8)       &  2.80   &  0         &  0.57      &  4.98 (-0.98, +0.97)  &     0.053	&   \nodata   \\
{\sc coup} 1443 & 5528$^{\dagger}$   & 0.5$^{\ddagger}$ & 2$^{\ddagger}$          & 3.74 (+4.17,-0.57)       &  13.6   &  \nodata   &  \nodata   &  0.26 (-0.26, +0.55)  &     0.18	&   0.14      \\
{\sc coup} 1568 & 5236               & 2.55             & 3.99                    & 0.53 (+1.62,-0.53)       &  48.1   &  \nodata   &  0.59      &  1.06 (-0.56, +0.60)  &    -0.040	&  -0.22      \\
{\sc coup} 1608 & 3724               & 0.48             & 1.76                    & 11.8 (+2.44,-11.8)       &  10.8   &  -1.3      &  0.93      &  0.25 (-0.25, +0.41)  &     0.43    	&   -1.2      \\
\enddata
\tablenotetext{a}{Taken from \citet{LAH:1997}, unless otherwise noted.}
\tablenotetext{b}{Taken from \citet{Favata:2005}. 
No uncertainty is quoted in cases where \citet{Favata:2005}
used only two points in fitting the flare $\log(T)$-$\log(n_e)$ decay slope.}
\tablenotetext{c}{Derived in this work.}
\tablecomments{
$\dagger$ Temperatures not in literature; assigned T$_{\rm eff}$ from best fit SED model. 
$\ddagger$ Fiducial masses and radii of 0.5 M$_{\sun}$ and 2 R$_{\sun}$ taken in cases for which there were no measurements 
available in the literature.
}
\end{deluxetable}
\end{landscape}

\clearpage

\begin{landscape}
\begin{deluxetable}{l|ccccc|cccc}
\tabletypesize{\scriptsize}
\tablecaption{\textit{Hubble} {\sc ACS} and {\sc WFI}  Fluxes \label{fluxes1}}
\tablewidth{0pt}
\tablehead{\colhead{Object}  & \colhead{4317\AA}    & \colhead{5359\AA}    & \colhead{6584\AA}\tablenotemark{(1)}    & \colhead{7693\AA}    & \colhead{9055\AA}    & \colhead{0.36$\mu$m}    & \colhead{0.44$\mu$m}    & \colhead{0.55$\mu$m}  & \colhead{0.83$\mu$m}    \\
           \colhead{Name}    & \colhead{Flux [mJy]} & \colhead{Flux [mJy]} & \colhead{Flux [mJy]} & \colhead{Flux [mJy]} & \colhead{Flux [mJy]} & \colhead{Flux [mJy]} & \colhead{Flux [mJy]} & \colhead{Flux [mJy]}  & \colhead{Flux [mJy]} }
\startdata
{\sc coup} 7             & \nodata              & \nodata              &     170      & \nodata              & \nodata              &     3.9$\pm$0.36     &     40.$\pm$3.7      &    83.0$\pm$7.6      &    290$\pm$27       \\
{\sc coup} 28            &    0.89$\pm$0.082    &     2.9$\pm$0.27     &     7.3      & \nodata              & \nodata              &   0.099$\pm$0.0091   &     1.1$\pm$0.10     &     2.7$\pm$0.25     &     16$\pm$1.5      \\
{\sc coup} 43            &    0.40$\pm$0.037    &     1.6$\pm$0.15     &     5.2      &     10.$\pm$0.95     &     20.$\pm$1.9      &   0.033$\pm$0.0031   &    0.42$\pm$0.039    &     1.2$\pm$0.11     &     15$\pm$1.4      \\
{\sc coup} 90            &   0.028$\pm$0.0026   &    0.15$\pm$0.014    &    0.69      &     1.7$\pm$0.15     &     4.0$\pm$0.37     & \nodata              &   0.038$\pm$0.0035   &    0.13$\pm$0.012    &    2.9$\pm$0.26     \\
{\sc coup} 141           &     7.9$\pm$0.73     &      16$\pm$1.4      & \nodata      & \nodata              & \nodata              &     1.3$\pm$0.12     & \nodata              & \nodata              &     66$\pm$6.1      \\
{\sc coup} 223           &   0.074$\pm$0.0068   &    0.32$\pm$0.029    &     1.6      & \nodata              & \nodata              & \nodata              &   0.095$\pm$0.0088   & \nodata              &    6.2$\pm$0.57     \\
{\sc coup} 262           &   0.011$\pm$0.0010   &   0.095$\pm$0.0088   &    0.57      &     2.0$\pm$0.18     &     6.7$\pm$0.61     & \nodata              & \nodata              &    0.11$\pm$0.010    &    4.4$\pm$0.41     \\
{\sc coup} 332           & \nodata              & \nodata              & \nodata      &   0.049$\pm$0.0045   &    0.33$\pm$0.030    & \nodata              & \nodata              & \nodata              &    \nodata          \\
{\sc coup} 342           &   0.008$\pm$0.00072  &   0.078$\pm$0.0072   &    0.52      &     1.9$\pm$0.17     &     6.3$\pm$0.58     & \nodata              & \nodata              &   0.099$\pm$0.0091   &    4.4$\pm$0.41     \\
{\sc coup} 454           &    0.11$\pm$0.010    &    0.64$\pm$0.059    &     2.6      &     6.8$\pm$0.62     &      16$\pm$1.5      & \nodata              &    0.16$\pm$0.014    &    0.63$\pm$0.058    &     13$\pm$1.2      \\
{\sc coup} 597           &     2.0$\pm$0.18     &     5.8$\pm$0.54     & \nodata      & \nodata              &      26$\pm$2.4      &    0.34$\pm$0.032    &     2.3$\pm$0.22     &     5.3$\pm$0.48     &     23$\pm$2.1      \\
{\sc coup} 649           &   0.019$\pm$0.0017   &    0.10$\pm$0.0093   &    0.38      &     1.5$\pm$0.14     &     3.4$\pm$0.31     & \nodata              & \nodata              & \nodata              &    2.7$\pm$0.25     \\
{\sc coup} 669           &     1.8$\pm$0.17     &     6.6$\pm$0.61     &      15      & \nodata              & \nodata              & \nodata              &     2.4$\pm$0.22     &     5.6$\pm$0.52     &     31$\pm$2.8      \\
{\sc coup} 720           & \nodata              & \nodata              &   0.070      &    0.31$\pm$0.029    &     1.5$\pm$0.14     & \nodata              & \nodata              & \nodata              &   0.99$\pm$0.091    \\
{\sc coup} 752           &    0.85$\pm$0.078    &     2.6$\pm$0.24     &     7.1      &     10.$\pm$0.92     &      16$\pm$1.4      &    0.18$\pm$0.016    &     1.3$\pm$0.12     &     3.0$\pm$0.27     &     16$\pm$1.5      \\
{\sc coup} 848           &   0.097$\pm$0.0089   &    0.37$\pm$0.034    &     1.5      &     3.3$\pm$0.30     &     6.9$\pm$0.63     & \nodata              & \nodata              &    0.46$\pm$0.042    &    5.4$\pm$0.50     \\
{\sc coup} 891           &   0.002$\pm$0.00022  &   0.046$\pm$0.0043   &    0.40      &     2.1$\pm$0.20     &     9.1$\pm$0.84     & \nodata              & \nodata              &   0.046$\pm$0.0042   &    5.7$\pm$0.53     \\
{\sc coup} 915           & \nodata              & \nodata              & \nodata      & \nodata              & \nodata              & \nodata              & \nodata              & \nodata              &   0.17$\pm$0.015    \\
{\sc coup} 960           &   0.022$\pm$0.0020   &   0.082$\pm$0.0075   &    0.40      &     1.6$\pm$0.14     &     4.3$\pm$0.40     & \nodata              &   0.029$\pm$0.0027   &   0.076$\pm$0.0070   &    3.1$\pm$0.28     \\
{\sc coup} 971           &     10.$\pm$0.96     &      21$\pm$2.0      &      43      & \nodata              & \nodata              &     1.6$\pm$0.15     &      11$\pm$1.0      & \nodata              &     64$\pm$5.9      \\
{\sc coup} 976           &   0.016$\pm$0.0015   &   0.080$\pm$0.0074   &    0.51      &     1.4$\pm$0.13     &     3.8$\pm$0.35     & \nodata              & \nodata              & \nodata              &    2.7$\pm$0.25     \\
{\sc coup} 997           &    0.60$\pm$0.055    &     2.1$\pm$0.19     &     5.9      &      11$\pm$0.98     &      18$\pm$1.7      & \nodata              &    0.77$\pm$0.071    &     2.0$\pm$0.18     &     14$\pm$1.3      \\
{\sc coup} 1040          & \nodata              & \nodata              & \nodata      & \nodata              & \nodata              & \nodata              & \nodata              & \nodata              &   0.10$\pm$0.0093   \\
{\sc coup} 1083          &    0.43$\pm$0.040    &     1.8$\pm$0.17     &     6.2      &      14$\pm$1.3      &      26$\pm$2.4      & \nodata              &    0.55$\pm$0.051    &     1.7$\pm$0.16     &     18$\pm$1.7      \\
{\sc coup} 1114          &    0.23$\pm$0.021    &     1.6$\pm$0.15     &     7.0      & \nodata              & \nodata              & \nodata              &    0.38$\pm$0.035    &     1.6$\pm$0.14     &     37$\pm$3.4      \\
{\sc coup} 1246          &   0.046$\pm$0.0043   &    0.22$\pm$0.020    &     1.2      &     1.3$\pm$0.12     &     5.2$\pm$0.48     & \nodata              &   0.059$\pm$0.0055   &    0.19$\pm$0.017    &    3.9$\pm$0.36     \\
{\sc coup} 1343          &   0.058$\pm$0.0054   &    0.31$\pm$0.028    &     1.7      &     2.8$\pm$0.26     &     6.7$\pm$0.62     & \nodata              &   0.082$\pm$0.0075   &    0.27$\pm$0.025    &    5.0$\pm$0.46     \\
{\sc coup} 1384          &     2.8$\pm$0.26     &     7.9$\pm$0.73     &      17      & \nodata              & \nodata              &    0.39$\pm$0.036    &     3.3$\pm$0.30     &     6.9$\pm$0.64     &    30.$\pm$2.7      \\
{\sc coup} 1410          &  0.0040$\pm$0.00037  &   0.022$\pm$0.0020   &    0.13      &    0.48$\pm$0.045    &     1.4$\pm$0.13     & \nodata              & \nodata              &   0.021$\pm$0.0019   &    1.0$\pm$0.093    \\
{\sc coup} 1440          &     1.9$\pm$0.18     &     6.2$\pm$0.57     &      15      & \nodata              &      24$\pm$2.2      &    0.21$\pm$0.019    &     2.5$\pm$0.23     &     6.0$\pm$0.56     &     31$\pm$2.8      \\
{\sc coup} 1568          & \nodata              &      85$\pm$7.8      & \nodata      & \nodata              & \nodata              &     8.4$\pm$0.77     &     50.$\pm$4.6      & \nodata              &    202$\pm$19       \\
{\sc coup} 1608          & \nodata              &     1.4$\pm$0.13     & \nodata      & \nodata              & \nodata              &    0.20$\pm$0.018    &    0.73$\pm$0.068    & \nodata              &     11$\pm$1.02     \\ \hline
Zero points        & 25.793\tablenotemark{(2)} & 25.744\tablenotemark{(2)} & 22.393\tablenotemark{(2)} & 25.291\tablenotemark{(2)} & 24.347\tablenotemark{(2)} & 1823 Jy\tablenotemark{(3)}& 4130 Jy\tablenotemark{(3)}& 3640 Jy\tablenotemark{(3)}& 2430 Jy\tablenotemark{(3)} \\
\enddata
\tablecomments{The first five data columns report {\it HST} fluxes, the last
four columns report WFI fluxes. Zeropoint fluxes used are reported in
the final row of the table.}
\tablenotetext{(1)}{For the purposes of fitting, the H$\alpha$ fluxes were given a wide (99\%) error in order to allow for variability commonly observed T Tauri stars.}
\tablenotetext{(2)}{The ACS data utilized were in the Vegamag system; these
values were used in the conversion to the ABmag system, for which flux
calculation is straightforward as outlined in the online ACS documentation
found here: \url{http://www.stsci.edu/hst/acs/analysis/zeropoints/}}
\tablenotetext{(3)}{Johnson-Cousins zero points; UBVI$_{C}$.}
\end{deluxetable}
\end{landscape}

\clearpage

\begin{landscape}
\begin{deluxetable}{l|cc|ccc|ccccc}
\tabletypesize{\scriptsize}
\tablecaption{$V$, $I$, {\sc 2MASS}, and \textit{Spitzer}  Fluxes \label{fluxes2}}
\tablewidth{0pt}
\tablehead{\colhead{Object}  & \colhead{0.55$\mu$m}    & \colhead{0.79$\mu$m}    & \colhead{1.235~$\mu$m} & \colhead{1.662~$\mu$m} & \colhead{2.159~$\mu$m} & \colhead{3.6~$\mu$m}  & \colhead{4.5~$\mu$m}  & \colhead{5.8~$\mu$m}  & \colhead{8.0~$\mu$m}  & \colhead{23.6~$\mu$m} \\
           \colhead{Name}    & \colhead{Flux [mJy]} & \colhead{Flux [mJy]} & \colhead{Flux [mJy]}  & \colhead{Flux [mJy]}  & \colhead{Flux [mJy]}  & \colhead{Flux [mJy]} & \colhead{Flux [mJy]} & \colhead{Flux [mJy]} & \colhead{Flux [mJy]} & \colhead{Flux\tablenotemark{a} [mJy]} }
\startdata
{\sc coup} 7             &     102$\pm$9.4      &     280$\pm$25       &    460$\pm$42         &    590$\pm$54         &     440$\pm$41        &     200$\pm$18      &     130$\pm$12      &      84$\pm$7.8     &     50.$\pm$4.6     & 10                   \\
{\sc coup} 28            &     3.8$\pm$0.35     &      17$\pm$1.6      &     39$\pm$3.6        &     47$\pm$4.4        &      41$\pm$3.8       &     20.$\pm$1.8     &      13$\pm$1.2     &     8.9$\pm$0.82    &     6.5$\pm$0.60    & 14                   \\
{\sc coup} 43            &     2.2$\pm$0.20     &      15$\pm$1.4      &     51$\pm$4.7        &     72$\pm$6.6        &      62$\pm$5.7       &      33$\pm$3.0     &      23$\pm$2.1     &      16$\pm$1.4     &      11$\pm$0.98    & \nodata \\
{\sc coup} 90            &   0.084$\pm$0.0078   &     1.8$\pm$0.16     &     14$\pm$1.2        &     27$\pm$2.5        &      27$\pm$2.5       &      16$\pm$1.5     &      11$\pm$1.0     &     6.6$\pm$0.60    &   \nodata           & \nodata \\
{\sc coup} 141           &      24$\pm$2.2      &      71$\pm$6.6      &    130$\pm$12         &    180$\pm$16         &     170$\pm$16        &     150$\pm$14      &     140$\pm$13      &     120$\pm$11      &     160$\pm$15      & \nodata \\
{\sc coup} 223           &    0.42$\pm$0.038    &     5.1$\pm$0.47     &     39$\pm$3.6        &     93$\pm$8.6        &     120$\pm$11        &     140$\pm$13      &     130$\pm$12      &     110$\pm$10.     &      91$\pm$8.4     & \nodata \\
{\sc coup} 262           &    0.31$\pm$0.028    &     2.7$\pm$0.25     &     35$\pm$3.2        &     96$\pm$8.8        &     130$\pm$12        &      98$\pm$9.0     &      68$\pm$6.3     &      39$\pm$3.6     &     \nodata         & \nodata              \\   
{\sc coup} 332           & \nodata              & \nodata              &    4.8$\pm$0.44       &     22$\pm$2.0        &     40.$\pm$3.7       &      64$\pm$5.9     &      69$\pm$6.3     &      87$\pm$8.0     &     150$\pm$14      & \nodata              \\
{\sc coup} 342           & \nodata              & \nodata              &     32$\pm$3.0        &     81$\pm$7.5        &      95$\pm$8.7       &      72$\pm$6.6     &     50.$\pm$4.6     &      63$\pm$5.8     & \nodata             & \nodata              \\   
{\sc coup} 454           &    0.66$\pm$0.061    &     10.$\pm$0.93     &     74$\pm$6.8        &    150$\pm$14         &     150$\pm$14        &     150$\pm$14      &     130$\pm$12      &     120$\pm$11      &     150$\pm$14      & \nodata              \\  
{\sc coup} 597           &     6.1$\pm$0.56     &      21$\pm$1.9      &     41$\pm$3.8        &     58$\pm$5.4        &      63$\pm$5.8       &      71$\pm$6.5     &      62$\pm$5.7     & \nodata             & \nodata             & \nodata              \\   
{\sc coup} 649           &   0.082$\pm$0.0075   &     1.6$\pm$0.15     &     14$\pm$1.3        &     29$\pm$2.6        &      31$\pm$2.8       & \nodata             &    \nodata          &  \nodata            &    \nodata          & \nodata              \\   
{\sc coup} 669           &     5.8$\pm$0.54     &      24$\pm$2.2      &     68$\pm$6.3        &     97$\pm$8.9        &      83$\pm$7.7       &     110$\pm$9.9     &    \nodata          &   \nodata           &    \nodata          & \nodata              \\   
{\sc coup} 720           & \nodata              & \nodata              &     13$\pm$1.2        &     43$\pm$4.0        &      65$\pm$6.0       &      83$\pm$7.7     &      70$\pm$6.5     &      58$\pm$5.4     &      57$\pm$5.2     & \nodata \\
{\sc coup} 752           &     3.4$\pm$0.31     &      12$\pm$1.1      &     31$\pm$2.8        &     40$\pm$3.7        &      34$\pm$3.1       &     20.$\pm$1.9     &      21$\pm$1.9     &      15$\pm$1.4     &     20.$\pm$1.8     & 11                   \\   
{\sc coup} 848           &    0.36$\pm$0.033    &     4.1$\pm$0.37     &     17$\pm$1.6        &     22$\pm$2.0        &     20.$\pm$1.9       &      \nodata        &     \nodata         &    \nodata          &    \nodata          & \nodata              \\      
{\sc coup} 891           &    0.10$\pm$0.0094   &     3.4$\pm$0.31     &     69$\pm$6.3        &    240$\pm$22         &     350$\pm$32        &     240$\pm$22      &     170$\pm$15      &     120$\pm$11      &      79$\pm$7.3     & 90                   \\   
{\sc coup} 915           & \nodata              & \nodata              &    5.6$\pm$0.52       &     34$\pm$3.1        &      68$\pm$6.3       &      78$\pm$7.2     &      67$\pm$6.2     &      56$\pm$5.1     &     \nodata         & \nodata \\
{\sc coup} 960           &   0.093$\pm$0.0086   &     2.1$\pm$0.19     &     11$\pm$1.0        &     13$\pm$1.2        &      11$\pm$1.0       &     6.6$\pm$0.61    &     4.7$\pm$0.43    &    \nodata          &    \nodata          & 36                   \\      
{\sc coup} 971           &      26$\pm$2.4      &      62$\pm$5.7      &    102$\pm$9.4        &    140$\pm$13         &      82$\pm$7.6       &      44$\pm$4.1     & \nodata             &    \nodata          & \nodata             & \nodata \\
{\sc coup} 976           & \nodata              & \nodata              &     13$\pm$1.2        &     24$\pm$2.2        &      26$\pm$2.4       &      24$\pm$2.2     &     20.$\pm$1.8     &    \nodata          &    \nodata          & \nodata              \\      
{\sc coup} 997           &     2.0$\pm$0.19     &      11$\pm$0.98     &     36$\pm$3.3        &     55$\pm$5.0        &     50.$\pm$4.6       &      39$\pm$3.6     &      27$\pm$2.5     & \nodata             & \nodata             & \nodata              \\      
{\sc coup} 1040          & \nodata              & \nodata              &    4.0$\pm$0.37       &     26$\pm$2.4        &      58$\pm$5.3       &      75$\pm$6.9     &      76$\pm$7.0     &     60.$\pm$5.6     &      56$\pm$5.2     & \nodata \\
{\sc coup} 1083          &     3.2$\pm$0.29     &      16$\pm$1.5      &     57$\pm$5.2        &     95$\pm$8.8        &      86$\pm$7.9       &      54$\pm$5.0     &      36$\pm$3.3     &    \nodata          & \nodata             & \nodata              \\      
{\sc coup} 1114          &     1.9$\pm$0.17     &      26$\pm$2.4      &    190$\pm$18         &    380$\pm$35         &     410$\pm$37        &     250$\pm$23      &     190$\pm$18      &     130$\pm$12      &    86$\pm$7.9       & \nodata \\
{\sc coup} 1246          &    0.28$\pm$0.025    &     3.2$\pm$0.29     &     15$\pm$1.3        &     24$\pm$2.2        &      28$\pm$2.5       &      18$\pm$1.7     &      17$\pm$1.5     &      21$\pm$1.9     &      42$\pm$3.9     & \nodata \\
{\sc coup} 1343          &    0.32$\pm$0.029    &     3.5$\pm$0.32     &     25$\pm$2.3        &     58$\pm$5.4        &      82$\pm$7.5       &      70$\pm$6.5     &      62$\pm$5.7     &      35$\pm$3.2     &     \nodata         & \nodata              \\      
{\sc coup} 1384          &     8.2$\pm$0.75     &      28$\pm$2.6      &     66$\pm$6.1        &     87$\pm$8.0        &      69$\pm$6.4       &      34$\pm$3.1     &      23$\pm$2.1     &      18$\pm$1.6     &     \nodata         & \nodata              \\      
{\sc coup} 1410          &    0.14$\pm$0.013    &    0.72$\pm$0.067    &    5.8$\pm$0.53       &     12$\pm$1.1        &      12$\pm$1.1       &     7.2$\pm$0.67    &     4.8$\pm$0.44    &    \nodata          & \nodata             & \nodata \\
{\sc coup} 1443          &     6.5$\pm$0.60     &      25$\pm$2.3      &     57$\pm$5.3        &     77$\pm$7.1        &      62$\pm$5.7       &      32$\pm$3.0     &     20.$\pm$1.9     &      18$\pm$1.7     &     \nodata         & \nodata              \\      
{\sc coup} 1568          &     109$\pm$10.      &     207$\pm$19       &    290$\pm$26         &    300$\pm$27         &     240$\pm$22        &     130$\pm$12      &    100.$\pm$9.6     &      97$\pm$8.9     &     130$\pm$11      & 156$\pm$14 \\
{\sc coup} 1608          & \nodata              & \nodata              &     26$\pm$2.4        &     39$\pm$3.6        &      46$\pm$4.2       &      45$\pm$4.1     &      46$\pm$4.3     &      34$\pm$3.2     &      43$\pm$3.9     & 119$\pm$16           \\ \hline
Zero points [Jy]   & 3640\tablenotemark{(1)}&  2490\tablenotemark{(2)}& 1594\tablenotemark{(3)}& 1024\tablenotemark{(3)}& 666.7\tablenotemark{(3)}& 280.9\tablenotemark{(4)}& 179.7\tablenotemark{(4)}& 115.\tablenotemark{(4)}& 64.13\tablenotemark{(4)}& 7.17\tablenotemark{(5)} \\
\enddata
\tablenotetext{a}{Fluxes reported without error are 3$\sigma$ upper limits.}
\tablenotetext{1.}{\citet{Bessell:1979}.}
\tablenotetext{2.}{\citet{Cousins:1976}.}
\tablenotetext{3.}{\citet{Cohen:2003}.}
\tablenotetext{4.}{Spitzer Science Center data calibration manual, accessible online at: \url{http://ssc.spitzer.caltech.edu/documents/cookbook/html/cookbook-node208.html}}
\tablenotetext{5.}{\citet{Engelbracht:2007}.}
\end{deluxetable}
\end{landscape}

\clearpage

\begin{table}
\caption{Reliability of Near-IR Excess as Tracer of 
Inner Disk Edge\label{foursquare}}
\begin{tabular}{|p{4cm}|p{4cm}|p{4cm}|}
\hline \nodata & R$_{\rm trunc} \lesssim$~R$_{\rm dust}$ & R$_{\rm trunc} >$~R$_{\rm dust}$ \\ \hline
$\Delta(K_{S}) \ge 0.3$ & 141\tablenotemark{a}, 223, 1608\tablenotemark{a}  & 1246\tablenotemark{b} \\ \hline
$\Delta(K_{S}) < 0.3$     & 332, 454, 720, 752, 976, 1040, 1384, 1568 & 7, 28, 43, 90, 262, 597, 669, 891, 960, 971, 1083, 1114, 1410, 1443 \\ \hline
\end{tabular}
\tablenotetext{a}{Object possesses \ion{Ca}{2} in emission (Table~\ref{stelparms})
a spectroscopic indicator of active accretion.}
\tablenotetext{b}{We consider the large \dik\ for COUP 1246 to be spurious (see discussion of this specific case in \S\ref{excessive}), so this object actually belongs in the lower right quadrant of the Table.}
\tablecomments{Category 4 objects have been omitted, as a 
clear determination of $R_{\rm trunc}$ could not be made; see \S\ref{criteria}.}
\end{table}

\begin{table}
\caption{Reliability of Near-UV Excess as Tracer of 
Inner Disk Edge\label{foursquare2}}
\begin{tabular}{|p{4cm}|p{4cm}|p{4cm}|}
\hline \nodata & R$_{\rm trunc} \lesssim$~R$_{\rm dust}$ & R$_{\rm trunc} >$~R$_{\rm dust}$ \\ \hline
$\Delta(U-V) \le -0.3$ & 141\tablenotemark{a}, 752, 1384, 1608\tablenotemark{a}  &  971 \\ \hline
$\Delta(U-V) > -0.3$   & 1568 & 7, 28, 43, 597, 1443 \\ \hline
\end{tabular}
\tablenotetext{a}{Object possesses \ion{Ca}{2} in emission (Table~\ref{stelparms})
a spectroscopic indicator of active accretion.}
\tablecomments{Category 4 objects have been omitted, as a 
clear determination of $R_{\rm trunc}$ could not be made; see \S\ref{criteria}.}
\end{table}

\begin{deluxetable}{cccc}
\tabletypesize{\scriptsize}
\tablecaption{Spectral Energy Distribution Result Summary\label{summary}}
\tablewidth{0pt}
\tablehead{\colhead{Category 1} & \colhead{Category 2} & \colhead{Category 3} & \colhead{Category 4}}
\startdata
332     &	141     &	7	&	223     \\
454     &	597     &	28     	&	342     \\
720     &	1040    &	43     	&	649	\\
752     &	1568    &	90     	&	848	\\
1384    &	\nodata	&	262   	&	915    	\\
1608    &	\nodata	&	669    	&	976    	\\
\nodata &	\nodata	&	891     &	997    	\\
\nodata	&	\nodata	&	960     &	1343   	\\
\nodata	&	\nodata	&	971	&	\nodata	\\
\nodata &	\nodata &	1083	&	\nodata \\
\nodata	&	\nodata	&	1114    &	\nodata	\\
\nodata	&	\nodata	&	1246    &	\nodata	\\
\nodata	&	\nodata	&	1410    &	\nodata	\\
\nodata	&	\nodata	&	1443    &	\nodata	\\
\enddata
\tablecomments{Categories as described in \S \ref{criteria} for all {\sc coup} sample objects.}
\end{deluxetable}


\clearpage

\clearpage

\clearpage

\clearpage

\clearpage

\clearpage

\bibliographystyle{apj}
\bibliography{refs}

\end{document}